\documentclass{aastex}
\usepackage{emulateapj5,apjfonts} 
\usepackage{graphicx}
\newcommand{\be}{\begin{equation}}
\newcommand{\ba}{\begin{eqnarray}}
\newcommand{\ee}{\end{equation}}
\newcommand{\ea}{\end{eqnarray}}  
\newcommand{\etal}{et al.\ }
\def\lesssim{\mathrel{\hbox{\rlap{\hbox{\lower4pt\hbox{$\sim$}}}\hbox{$<$}}}}
\def\gtrsim{\mathrel{\hbox{\rlap{\hbox{\lower4pt\hbox{$\sim$}}}\hbox{$>$}}}}

\def\gtsima{$\; \buildrel > \over \sim \;$}
\def\ltsima{$\; \buildrel < \over \sim \;$}
\def\gsim{\lower.5ex\hbox{\gtsima}}
\def\lsim{\lower.5ex\hbox{\ltsima}}
\def\simgt{\lower.5ex\hbox{\gtsima}}
\def\simlt{\lower.5ex\hbox{\ltsima}}
\def\simpr{\lower.5ex\hbox{\prosima}}

\def\msun{{M_\odot}}

\def\simless{\mathbin{\lower 3pt\hbox
   {$\rlap{\raise 5pt\hbox{$\char'074$}}\mathchar''7218$}}}   
\def\simgreat{\mathbin{\lower 3pt\hbox
   {$\rlap{\raise 5pt\hbox{$\char'076$}}\mathchar''7218$}}}   

\begin{document}

\title{The Impact of Small-Scale Structure on Cosmological Ionization Fronts
  and Reionization}
\author{Ilian~T.~Iliev$^1$, Evan Scannapieco,$^2$ and Paul~R.~Shapiro$^3$}

\altaffiltext{1}{Canadian Institute for Theoretical Astrophysics, University
  of Toronto, 60 St. George Street, Toronto, ON M5S 3H8, Canada}
\altaffiltext{2}{Kavli Institute for Theoretical Physics, Kohn Hall, UC Santa
  Barbara, Santa Barbara, CA 93106}

\altaffiltext{3}{Department of Astronomy, University of Texas, Austin, TX
  78712-1083}
\submitted{Accepted by ApJ}
\label{firstpage}

\begin{abstract}
  
The propagation of cosmological ionization fronts during the
reionization of the universe is strongly influenced by small-scale
gas inhomogeneities due to structure formation.  These inhomogeneities
include both collapsed minihalos, which are generally self-shielding,
and lower-density structures, which  are not.  The minihalos are dense
and sufficiently optically-thick to trap intergalactic ionization fronts, 
blocking their path and robbing them of ionizing
photons until the minihalo gas is expelled as an evaporative wind.  The
lower-density structures do not trap these  fronts, but they can slow
them down by increasing the overall recombination rate in the
intergalactic medium (IGM). In this paper we study the effects of both
types of inhomogeneities, including nonlinear clustering effects,
and we find that both IGM clumping and collapsed minihalos have
significant yet qualitatively different impacts on reionization. While
the number density of minihalos on average increases strongly with
time, the density of minihalos {\em inside H~II regions around
ionizing sources} is largely constant.  Thus the impact of minihalos is
essentially to decrease the number of ionizing photons  available to
the IGM at all epochs, which is equivalent to a reduction in the
luminosity of each source.  On the other hand, the effect of IGM
clumping increases strongly with time, slowing down reionization and
extending it.  Thus while the impact of minihalos is largely
degenerate with the unknown source efficiency, IGM clumping can help
significantly in reconciling the recent observations of cosmic
microwave background polarization with  quasar absorption spectra at
$z \sim 6$, which together point to an early but extended reionization
epoch.

\end{abstract}


\section{Introduction}

Recent polarization observations of the cosmic microwave background by
the {\em Wilkinson Microwave  Anisotropy  Probe} (WMAP) imply that
reionization was fairly advanced at  $z_{\rm re} \sim 15 $ \citep{Ketal03}.  
This came as a surprise.  The prior detection of the
Gunn-Peterson effect in the spectra of  high-redshift quasars  had
suggested that reionization was only just ending at $z \sim 6$ 
\citep{W03,F04}. That was consistent with predictions of
the most accurate numerical simulations in the current $\Lambda$CDM
paradigm, which had all   predicted this transition  at $z_{\rm
re}\lesssim8-10$  \citep{CFGJ00,G00a,RNAS02,CSW03}.  Despite many
poorly understood details concerning the star formation rate, the
escape fraction of ionizing radiation, and the differences in
numerical treatments of reionization, $z_{\rm re} \sim 15$ had seemed
unlikely, and such an extended period of reionization, impossible.

Now the race is on to reconcile the early onset of reionization
suggested by WMAP with the high-redshift Gunn-Peterson effect, which
implies  neighboring ionized patches finally grew to overlap at
$z\sim6$ \citep{HH03,C03,WL03,CFW03}. One suggestion is that the
universe had two reionization epochs but recombined in between
\citep{C03}, yet this ignores the unavoidable spread in redshifts
intrinsic to any such IGM transition \citep{SSF03,BL04,FL05}.  Other
suggestions involve fine-tuning the ionizing photon emissivity for
different source halo masses, the escape fraction, and the (possibly
metalicity-dependent) Initial Mass Function (IMF), in ways intended
to accelerate early ionization, to build up a large enough $\tau_{\rm
es}$, but slow down late ionization, to delay the final overlap until
$z \approx 6$ \citep{WL03,CFW03}.  Finally, several authors have
explored the possibility of early partial reionization due to a
decaying particle \citep{CK04,HH04}, complemented by later full
reionization from astrophysical sources.

\begin{figure*}
  \includegraphics[width=3.2in]{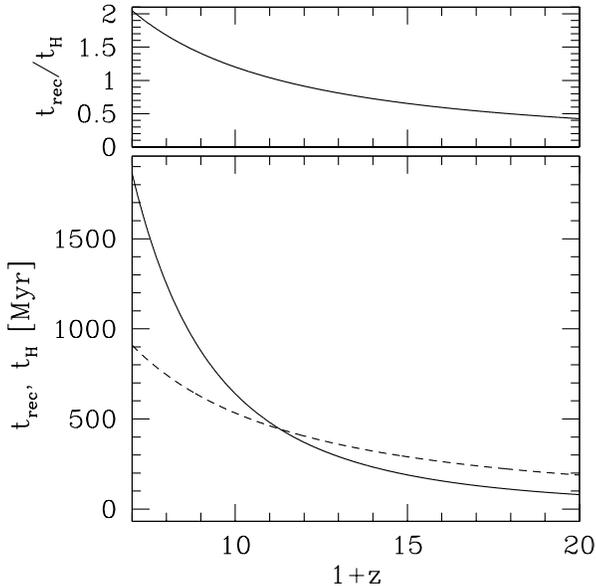}
\caption{Timescales. Hubble time $t_H$ (dashed line) and recombination 
time $t_{\rm rec}=(\alpha_Bn_H)^{-1}$ at the mean IGM density (solid line) 
vs.  redshift $z$ (lower panel) and the ratio of these timescales (top panel).}
\label{timescales}
\end{figure*}

The role of small-scale inhomogeneities as sinks of ionizing photons
has mostly been ignored in this context. Nevertheless, over a large
range of redshifts, the recombination time $t_{\rm rec}$ at the mean
IGM density is on the order of the corresponding Hubble time, as
illustrated in Figure~\ref{timescales}.  Thus the absorption of
ionizing photons during reionization happens predominantly in
overdense regions.  In hierarchical models like Cold Dark Matter (CDM),
the smallest structures are the first to collapse gravitationally and
dominate the photon consumption both during the ionization of a region
and afterwards, while balancing recombinations.

When the first sources turned on, they ionized the neutral, opaque IGM
around them by propagating weak R-type ionization fronts
(I-fronts). This type of  front moves outward supersonically with
respect to both the neutral gas in front of it and the ionized gas
behind it, so it races ahead of the hydrodynamical response of the
IGM.  This process was first described by \citet{S86}
and \citet{SG87},
who solved analytically for the time-varying radius of a spherical
I-front surrounding a point source in the expanding IGM and then used
this solution to  determine when H~II regions would grow to the point
of overlap, thereby completing  reionization.  In this study the
effect of density inhomogeneity on the motion of the I-front was
described by a mean gas clumping factor 
$C \equiv \langle n^2\rangle/\langle n\rangle^2$.  A
clumpy gas has $C > 1$, which causes the ionized gas to recombine more
frequently, increasing the opacity of the H~II region  to ionizing
photons, which reduces the flux reaching the I-front and slows it
down.  This approach has formed the basis  for many more recent
semi-analytical treatments of reionization (e.g.\ Haiman \&
Holder 2003; Wyithe \& Loeb 2003; Venkatesan,  Tumlinson, \& Shull
2003). The idea that reionization proceeded by the propagation of
weak,  R-type I-fronts which move too fast to be affected by the gas
dynamical disturbance  they create is also the basis for most of the
{\em numerical} simulations of reionization carried out to date (e.g.\
Razoumov \etal 2002; Ciardi, Ferrara, \& White, 2003;  Sokasian \etal
2004).  In particular, all numerical studies that add radiative
transfer to a pre-computed inhomogeneous cosmological density field
(i.e. the ``static'' limit) are assuming that there is no significant
back  reaction on the gas\footnote{An exception to this is the code 
developed in \citet{G00a} and \citet{RGS02}, which combines an
approximate treatment of radiative transfer with numerical 
cosmological gas dynamics.}.

The assumption of either a mean clumping factor or the static limit
to model the effect of density inhomogeneity on 
cosmological I-fronts is not correct even on
average, however, unless the clumps are either  optically thin or
absorb only a small fraction of the ionizing flux.  If a clump is
self-shielding, then the I-front that encounters it will not remain a
weak R-type front if the size of the clump is larger  than its Str\"omgren 
length (i.e. the length of a column of gas within which the
unshielded  arrival rate of ionizing photons just balances the
recombination rate).  In that case the denser gas of the clump must
slow the I-front down enough that the disturbed gas inside the clump
catches up to the I-front and affects its progress.  This transforms
the I-front from supersonic, R-type, to subsonic, D-type and ``traps''
the I-front inside the clump \citep{SIR04}.   If the
clump is gravitationally bound before the arrival of the I-front, then
the I-front will expel the gas from the clump as a supersonic
evaporative wind,  as long as the clump cannot bind photoionized gas
with $T \geq 10^4$ K.

The impact of small-scale inhomogeneities on the global I-fronts that
reionized  the universe depended upon the relative importance of
unshielded and shielded  overdense regions and their sizes, densities,
and abundances.  These IGM inhomogeneities can be divided into two
major types, both of which have been modeled only crudely in the
majority of reionization studies.  Pre-virialized objects, such as
filaments and still-collapsing halos, are usually described in terms
of  the mean clumping factor described above.  Current semi-analytical
models of reionization either assume a constant clumping factor
\citep{C03,HH03,TVS04}, a clumping factor derived from linear theory
\citep{MHR00,CFO03,WL03}, or ignore clumping altogether (i.e. assume
$C=1$) \citep{OM04}. In practice all these approaches are
over-simplified since the clumping factor of the IGM gas is dominated
by the highly-overdense nonlinear regions and evolves strongly with
redshift.

Modeling of virialized inhomogeneities in previous studies has been
even more approximate.  An important dividing line that separates two
distinct populations  of virialized halos is that defined by the
virial temperature,  $T_{\rm vir} = 10^4 K$.  In order for stars to
form inside halos, the gas must cool below the virial temperature to
become self-gravitating and gravitationally unstable.  Radiative
cooling in a purely atomic gas of primordial composition is
ineffective below $10^4 K$, however, so ``minihalos''-- halos in the
mass range  $10^4 \msun \lesssim M \lesssim 10^8 \msun$, with virial
temperatures below $10^4$ K  -- are only able to form stars by forming
H$_2$ molecules, which have the potential to cool the gas below the
virial temperature, by rotational-vibrational line excitations.  The
H$_2$ that forms in minihalos, though, is easily dissociated  by UV
photons in the Lyman-Werner bands between 11.2 and 13.6 eV, which are
produced in abundance by the first stars, long before the ionizing
background from such stars is able to reionize a  significant fraction
of the universe \citep{HRL97,HAR00,CFGJ00}. Thus a generic prediction
of current structure formation models is a large population of
minihalos that are unable to cool and form stars.

In that case, the dominant source of photons for reionization would 
have been the more massive halos (i.e. $M \gtrsim 10^8 \msun$)
with $T_{\rm vir} \geq
10^4 K$, in which atomic line cooling is efficient enough to enable
stars to form.  From the point of view of such a source halo, all
lines of sight will intersect a minihalo at a distance less than the
mean spacing between sources \citep{HAM01,S01,SIRM03,SIR04}.  Thus,
the intergalactic I-fronts must have found their paths blocked by
minihalos in every direction, which   trapped the fronts until the
minihalos were evaporated.

In \citet{SIR04} and \citet{ISR04},
we used high-resolution numerical gas dynamical simulations with
radiative  transfer to study the encounter between an intergalactic
I-front and  a minihalo in detail, for a wide range of conditions
expected during reionization.  These results yielded the number of
ionizing photons absorbed per minihalo atom during the time between
the arrival of the I-front and the evaporation of the minihalo gas,
$\xi$, as a function of the minihalo mass, source flux level and
spectrum, and the redshift of the encounter.  This is a fundamental  
ingredient  we will need here to determine how the presence of
minihalos affected global reionization.

This trapping of intergalactic I-fronts by minihalos, combined with the
increased recombination rate inside already ionized regions due to
small-scale clumping outside the minihalos, may help to explain how
reionization could have started early and ended late.  As the global
I-fronts advanced into fresh neutral regions, they generally
encountered minihalos that formed at the {\it unfiltered} (i.e. not
affected by any radiation feedback) rate of the universe without
reionization. The mass fraction collapsed into minihalos in such
regions grew over time, from 8\% to 24\% to 31\% from $z=15$ to 9 to
6, so the average number of extra photons consumed per atom by
photoevaporation must also have increased with time.  This may have enabled
minihalos to slow the advance of the global I-fronts, with increasing
effect toward late times.  Reionization simulations by \citet{CFW03}
for example, which neglected minihalos, found that if one assumes a
high escape fraction of ionizing photons from the source halos, then
the large value of electron scattering optical depth, $\tau_{\rm es}$
observed by WMAP can be achieved by the first stars in galaxies with
mass $M \gtrsim 10^9M_\odot$ as sources.  However, in this
case, reionization is completed far too early.  Minihalos may have the
potential to reconcile this discrepancy, increasing the duration of
the epoch of reionization and allowing for a similar high value of
$\tau_{\rm es}$,  while postponing the redshift of
overlap.

In this paper, we consider the impact of both minihalos and more
general IGM clumping in detail, and attempt to quantify their effects
on the duration of the reionization epoch.  Driven by measurements of
the cosmic microwave background, the number abundance of galaxy
clusters, and high redshift supernova distance estimates 
\citep{Setal03,ECF96,Petal99} we focus
our attention on the $\Lambda$CDM
cosmological model with parameters
$h=0.7$, $\Omega_0$ = 0.3, $\Omega_\Lambda$ = 0.7, $\Omega_b = 0.05$,
$\sigma_8 = 0.87$, and $n_p=1$, where $\Omega_0$, $\Omega_\Lambda$, and
$\Omega_b$ are the total matter, vacuum, and baryonic densities in
units of the critical density ($\rho_{\rm crit}$), 
$\sigma_8^2$ is the variance of linear
fluctuations filtered on the $8 h^{-1}{\rm Mpc}$ scale, and $n_p$ is the 
index of the primordial power spectrum. The \citet{EH99} transfer 
function is used throughout.

The structure of this work is as follows.  In \S\ref{single_sect}  we
generalize the approach of \citet{SG87} to account for
the effect of minihalo evaporation on the time-varying radius of a
spherical I-front.  This will require us to calculate the
statistically biased abundance of minihalos at the location of the
front and incorporate the simulation results for the ionizing photon
consumption rates per minihalo.  In \S\ref{global_sect}  we model the
global progress of reionization by summing the results from Section 2
over a  statistical distribution of source halos, leading to the
eventual overlap of neighboring  H II regions and the completion of
reionization. Our results and conclusions are given in
\S~\ref{summary_sect}.

\section{The Propagation of A Cosmological Ionization Front about 
a Single Source}
\label{single_sect}

\subsection{Cosmological Ionization Fronts in a Clumpy IGM}

When a source of ionizing radiation turns on in the expanding, neutral
IGM, a weak, R-type I-front propagates outward. If the IGM were
static, this front would decelerate continuously from the moment of
turn on, until, within a time comparable to the recombination time,
it almost reached the size of the Str\"omgren sphere.
This Str\"omgren sphere is
just large enough that  the total recombination rate of ionized atoms
inside it equals the ionizing photon luminosity of the central
source.  At this point the I-front drops to the R-critical speed of
twice the sound speed of the ionized gas, and the front transforms
from R-type to D-type, preceded by a shock.  Thereafter the I-front is
affected by the dynamical  response of the IGM.

This is not the case, however, in the expanding, average IGM. 
\citet{SG87} showed that, while it is formally possible to 
define an ``instantaneous'' Str\"omgren radius (which grows in time in
proportion to the cosmic scale factor), the actual I-front generally
does not reach this radius.  Instead, the I-front remains a weak
R-type front as long as  the source continues to shine, and 
it would not be correct, therefore, to describe the cosmological H~II 
region as a Str\"omgren sphere, a misnomer which  unfortunately 
appears in the literature of reionization.  

We shall follow the approach of \citet{SG87}, in which
the H~II region is bounded by an I-front whose speed is determined by
the I-front continuity jump condition, which balances the outward flux
of ionizing photons against the inward flux of newly created ions.
The flux that reaches the front will be determined by solving the
equation of transfer between the source and the front.  We will assume
that the IGM is spherically symmetric outside the source.  For
simplicity  the I-fronts are taken to be ``sharp'', i.e. the width of
the transition between the ionized region inside and the neutral
region outside the front is small compared to its radius. The
actual width is comparable to the absorption mean free path
on the neutral side. This assumption of small mean free path
is generally a good approximation for a
``soft'' Population II (Pop.~II) 
stellar spectrum where most ionizing photons have
energies near the ionization threshold of hydrogen, for which the
absorption cross-section due to neutral hydrogen is large. However,
I-fronts are somewhat wider for ``hard'' spectra like those expected
for massive Population  III (Pop.~III) stars 
and the power-law spectra of QSOs. In
these cases a larger  fraction of the ionizing photons are at higher
energies, corresponding to lower ionization cross-sections of neutral
hydrogen and helium, and thus our  approximation of sharp I-fronts is
less accurate.

Adopting this picture, we consider an ionizing source emitting
$\dot{N}_{\gamma}$ ionizing photons per unit time.  We define the comoving 
radius of
the ionized region as $r_I(t)$ and its comoving volume as $V_I=4\pi r_I^3/3$.
We are interested in H~II regions that at all times are much smaller than the
scale of the current horizon.  The jump condition across the I-front is given
by a balance of the flux of neutral atoms and photoionizing photons. In the
frame of the front, it can be written as 
\be 
n_{H,1} u_1 =\beta_i^{-1}F,
\label{jump}
\ee 
where $n_{H,1}$ is the undisturbed hydrogen number density (in proper 
coordinates) on the neutral side of the front,
$u_1=a(dr_I/dt)$ is the I-front peculiar velocity, $a=1/(1+z)$ is the scale
factor, and $\beta_i$ is the number of ionizing photons absorbed 
to create each ionized H atom that emerges on the ionized side.
In the absence of minihalos, $\beta_i= \chi_{\rm eff} \equiv 1+p A({\rm He})$,
which corrects for the presence
of helium, with $p=0,1$ or 2 if He is mostly neutral, singly ionized or doubly
ionized after the passage through the front,
and $A({\rm He})=0.08$ is the He abundance by number with respect to hydrogen.
Finally, $F$ is the number flux of ionizing photons at the current position of
the I-front, 
\be 
F=\frac{S(r_I,t)}{4\pi a^2r_I^2},
\label{flux0}
\ee 
where $S(r,t)$ is the number of photons emitted by the central source
which pass through a sphere of comoving radius $r$ per unit time,
given at $r=r_I$ as
\be 
S(r_I,t)=\dot{N}_{\gamma}-\frac{4\pi}3
r_I^3a^{-3}({n_H}^0)^2 C\alpha_B\chi_{\rm eff}, 
\label{eq:SrIt}
\ee 
i.e. the number of photons
emitted by the source per unit time minus the number of recombinations in the
current H~II region volume. Here $C$ is the volume-averaged clumping factor,
$\alpha_B = 2.6 \times 10^{-13}$ cm$^3$ s$^{-1}$ is the case B recombination
coefficient for hydrogen at $10^4 K$ and $n_{\rm H}^0$ is the comoving 
number density of hydrogen in present units, $1.87 \times 10^{-7}
(\Omega_b h^2/0.022)$ cm$^{-3}.$ In
all cases, it is safe to assume that the H~II regions are
cosmologically small, and hence no ionizing photons are lost to
redshifting below the hydrogen ionization threshold.

Combining equations~(\ref{jump}) - (\ref{eq:SrIt}), the evolution of
the comoving volume of the ionized region $V_I$ is given by 
\be
\frac{dV_I}{dt} \equiv 4 \pi r_I^2 \frac{dr_I}{dt} \equiv \frac{1}{\chi_{\rm
    eff}n_H^0} \dot N_\gamma - \alpha_B \, C \, (1+z)^3 \, n_H^0 \, V_I.
\label{eq:paul2}
\ee 
Defining 
\be 
V_{\rm S,i}=\frac{4\pi r_{\rm S,i}^3}{3}
=\frac{\dot{N}_\gamma}{\chi_{\rm eff}\alpha_BC(n_{\rm H}^0)^2}, \ee we can
write equation~(\ref{eq:paul2}) in dimensionless form 
\be
\frac{dy}{dx}=1-y(1+z)^3,
\label{nondim}
\ee 
where $y\equiv V_I/V_{\rm S,i}=(r_I/r_{\rm S,i})^3$, $x\equiv t/t_1$ and
$t_1=1/(\alpha_B Cn_{\rm H}^0)$ is the recombination time of the mean IGM at
present \citep{SG87}.

If we define $d\tau=dx/a^3=(1+z)^3dx$, equation~(\ref{nondim}) becomes \be
\frac{dy}{d\tau}=(1+z)^{-3}-y,
\label{nondim2}
\ee for which a formal solution is \be
y(t)=e^{-\tau(t)}\int_{\tau(t_i)}^{\tau(t)}d\tau'\frac{e^{\tau'}}{[1+z(\tau')]^{3}},
\label{yt}
\ee where $t_i$ is the time of source turn-on and for the flat, $\Lambda$CDM
model \be \frac{d\tau}{dz}=-\frac{(1+z)^2}{H_0t_1[\Omega_0(1+z)^3+
  \Omega_\Lambda]^{1/2}}.
\label{tau_z}
\ee Equation~(\ref{tau_z}) has a solution 
\be
\tau(z)=\kappa\left\{1-\left[\Omega_0(1+z)^3+ \Omega_\Lambda
  \right]^{1/2}\right\},
\label{tau_exact}
\ee 
where $\kappa\equiv\frac{2}{3H_0t_1\Omega_0}$ and the arbitrary constant of
integration is chosen so that $\tau(z=0)=0.$ At high redshift, before and
during reionization, we have $\Omega_0>>\Omega_\Lambda /(1+z)^3$ and the
solution~(\ref{tau_exact}) becomes 
\be
\tau(z)=\kappa[1-\Omega_0^{1/2}(1+z)^{3/2}].  
\ee 
In this limit
equation~(\ref{yt}) simplifies to 
\be 
y(t)=\Omega_0\kappa^2e^{-\tau(t)}
\int_{\tau(t_i)}^{\tau(t)}d\tau'\frac{e^{\tau'}}{(\kappa-\tau')^2}.
\label{ylim}
\ee 
Equation~(\ref{ylim}) has an exact analytical solution given by 
\be
y=\frac{\eta}{(1+z_i)^3}e^{\eta t_i/t}\left[\frac{t}{t_i}
  Ei(2,\eta\frac{t_i}{t})-Ei(2,\eta)\right],
\label{y_exact}
\ee 
where $\eta\equiv2(1+z_i)^{3/2}/(3H_0t_1\Omega_0^{1/2})$ and
$Ei(2,x)\equiv\int_1^\infty \frac{e^{-xt}}{t^2}dt$ is the Exponential
integral of second order. The solution in equation~(\ref{y_exact})
reduces  to the one in  equation~(10a) of \citet{SG87} for a flat,
matter-dominated universe with $\Omega_0=1$, $t=2/(3H),$ and thus
$\eta=(1+z_i)^3t_i/t_1.$

\subsection{The Average Effect of Minihalo Evaporation on Cosmological
  Ionization Front Propagation}

Having outlined a formalism to describe the expansion of ionization fronts 
in a $\Lambda$CDM cosmology, we next address the question of absorption by
minihalos.  As described in \S 1, when an intergalactic I-front
encounters an individual minihalo, it is trapped until the minihalo gas 
is evaporated.  For every minihalo atom, this process consumes $\xi$
ionizing photons.  Suppose we consider the average effect of this process
on the global I-front which moves through a medium comprised of minihalos 
embedded in the IGM.  The average speed throughout this compound medium will
be given by a modified I-front continuity jump condition which takes account 
of the additional photon consumption due to minihalos.  In particular,
the quantity $\beta_i$ in eq.\ (\ref{jump}) should now be
replaced by the following
\be
 \beta_i \equiv (1-f_{\rm coll})\chi_{\rm eff}+
[1+A({\rm He})]f_{\rm coll,MH}\bar\xi,
\label{eq:betai}
\ee
where $f_{\rm coll}$ is the total collapsed baryon fraction (i.e.
over {\em all} halo masses) and $f_{\rm coll, MH}$ is the collapsed
fraction of just the minihalos. Finally, if $\xi$ is the the number 
of ionizing photons  consumed per minihalo
atom in the encounter between the  intergalactic I-front and an
individual minihalo, then $\bar \xi$ is the appropriate average over the
distribution of minihalos at the instantaneous location of the global
I-front.
Inserting eq.\ (\ref{eq:betai}) into eq.\ (\ref{jump})
then yields
\be 
n_{H,1}
u_1=\frac{F}{(1-f_{\rm coll})\chi_{\rm eff}+
[1+A({\rm He})]f_{\rm coll,MH}\bar\xi},
\label{jump1}
\ee 
where $n_{H,1}$ refers to the total H atom density, including both 
the IGM and all halos.  

As usual the flux, $F$, in this I-front jump condition is determined by
integrating the equation of transfer over the ionized region 
between the source halo and the I-front.  By definition,
the minihalos originally inside this region
do not affect this integration, however, since they will already
have been evaporated by the passage of the global I-front,
thereby returning their atoms to the IGM inside the H~II region.
We assume, for simplicity, that the evaporated minihalo gas shares the 
mean clumping factor the IGM into which it is mixed.  In that case,
the flux at the I-front is given by 
\be 
F=\frac{{\dot{N}_\gamma} - \alpha_B \, C \, (1+z)^3 \,
  (n_H^0)^2 {\chi_{\rm eff}}\, (V_I-V_0)}{4\pi a^2r_I^2},
\label{flux}
\ee 
where we have been careful now to start our integration of the transfer 
equation from the Lagrangian volume of the source halo (i.e. a volume 
which, when multiplied by the mean density, gives the mass of the source halo).

Combining equations (\ref{jump1}) and (\ref{flux}), the evolution of 
the comoving volume of the H~II region in the presence of minihalos is 
then given by 
\ba 
\frac{dV_I}{dt} =
\frac{1}{\left[(1-f_{\rm coll})\chi_{\rm eff}+[1+A({\rm He})]f_{\rm coll,MH}
    \bar\xi \right]}\nonumber \\
\times\left[\frac{\dot{N}_\gamma}{n_H^0} - \alpha_B \, C \, (1+z)^3 \, n_H^0
  \chi_{\rm eff}\, (V_I-V_0)\right],
\label{eq:meanscreen}
\ea 
where initially $V_I = V_0.$   We note that the solutions in \S 2.1 in
the absence of minihalos, including the exact analytical solution for 
$V_I(t)$ and $r_I(t)$ in
equation\ (\ref{y_exact}), will be valid here 
in the presence of minihalos as well, if 
$\beta_i$ in equation (\ref{eq:betai}) 
and the clumping factor $C$ are constants,  and $t_1$ is
redefined as 
\be
t_1 \equiv \left(\alpha_B C n^0_H \frac{\chi_{\rm eff}}{\beta_i}\right)^{-1}.
\ee
In general, both $\beta_i$ and $C$ will not be constant, but nevertheless 
this limit provides a useful check and some insight, as well shall
see below.

We adopt a simple model to account for infall when computing the flux.
The radial coordinate, $r_I$, adopted in our formalism above is essentially a
Lagrangian one, in which we have assumed that the local mean density of the
gas around the source is equal to the cosmic mean IGM density.  
In reality, all massive sources are found in overdense
regions, due to the gravitational influence on the surrounding gas. We estimate
the map between the Lagrangian and Eulerian comoving radii using a simple
``top-hat'' picture, which is a simplified version of the model described in
\citet{B04}.

We compute the cross-correlation between a sphere of Lagrangian radius $r_I$
and spherical perturbation of mass equal to the source halo mass
$M_s = (4 \pi/3) r_0^3 \Omega_0 \rho_{\rm crit}$ as
\be 
\sigma^2(r_0,r_I) \equiv \frac{1}{2 \pi} \int_0^\infty k^2 dk P(k) W(k
r_0) W(k r_I), 
\ee 
where $P(k)$ is the initial matter power spectrum, linearly extrapolated
to the present, and  $W(x)$ is the spherical top-hat window function,
defined in Fourier space as $W(x)\equiv 3 \left[\frac{\sin (x)}{x^3} -
  {\cos(x)}{x^2} \right]$, where $r_0$ is the Lagrangian radius of the source
itself.  Defining $\sigma^2(r_0)\equiv \sigma^2(r_0,r_0)$, the expected value
of the linear overdensity of a sphere of Lagrangian radius $r_I$ about the
source is then given by 
\be 
\bar \delta_L(r_I) =\frac{1.69 D(z)}{D(z_s)}
\frac{\sigma^2(r_0,r_I)}{\sigma^2(r_0)}+1.  
\ee 
Here we use $\bar \delta_L$ to
denote the average density within the sphere, $D(z)$ is the linear growth
factor, $z_s$ is the collapse redshift for the halo,
and the ``$+1$'' appears since we are defining the overdensity as
$\rho/\bar \rho$ instead of $\rho/\bar \rho -1$. For simplicity, we assume
$z=z_s$, i.e. that the growth of the mass after it collapses due to 
secondary infall, is small.

The linear overdensity $\bar \delta_L$ can be related to the corresponding
nonlinear overdensity, $\bar \delta$, by the standard spherical collapse 
model \citep{P80}. In this case, both quantities can be expressed
parametrically, in terms of a ``collapse parameter'' $\theta$ as 
\be 
\bar
\delta = \frac{9}{2} \frac{(\theta - {\rm sin} \, \theta )^2} {(1 - {\rm cos}
  \, \theta)^3},
\label{eq:nL}
\ee and \be \bar \delta_L = \frac{3}{5}\left(\frac{3}{4}\right)^{2/3} (\theta
- {\rm sin} \, \theta)^{2/3} + 1.  \ee These equations define the relationship
between the Eulerian and Lagrangian comoving radii as \be r_{I,E} = r_I \bar
\delta(r_I,r_0)^{-1/3},
\label{eq:radius}
\ee 
since $\bar \delta$ given by equation ($\ref{eq:nL}$) above is the average
overdensity within an Eulerian sphere of Lagrangian radius $r_I$.

\begin{figure*}
  \includegraphics[width=3.2in]{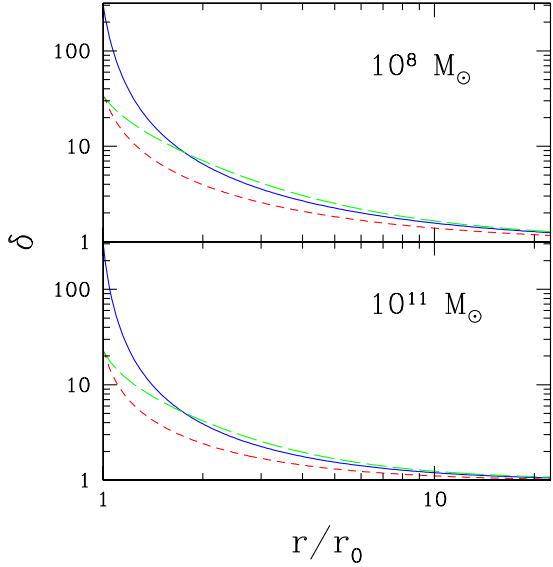}
\caption{Overdensites around a source of a given mass, as labeled,
versus Lagrangian distance from the center of the source 
(in units of the Lagrangian radius of the halo):
the mean $\bar \delta$ is solid (colored blue in electronic edition),
  the density at the boundary, $\delta$, is short-dashed (colored red in electronic edition), and
  $\delta_{\rm clump} $ is long-dashed (colored green in electronic edition). (See the electronic 
edition of the Journal for the color version of this figure.)}
\label{overdensities}
\end{figure*}

In the presence of infall, equation (\ref{eq:meanscreen}) becomes 
\ba
\frac{dV_I}{dt} =
\frac{1}{\left[(1-f_{\rm coll})\chi_{\rm eff}+[1+A({\rm He})]f_{\rm coll,MH}\bar\xi \right]}\nonumber \\
\times\left[\frac{\dot{N}_\gamma}{n_H^0} - \alpha_B \, C \, (1+z)^3 \,
  n_H^0\chi_{\rm eff} \,\int_{V_{0,E}}^{V_{I,E}} dV_{I,E}'
  \delta(V_{I,E}')^2\right],
\label{eq:meanscreen_infall}
\ea 
where now $\delta$ is not the average $\delta$ within the sphere, but
rather $\delta$ at the boundary, and $V_{I,E} \equiv V_I \bar \delta^{-1}$ is
the Eulerian volume.  We compute $\delta$ as follows. The comoving volume
satisfies: 
\be 
\Delta V_{I,E} \delta(V_{I,E}) + V_{I,E} \, \bar
\delta(V_{I,E}) = [V_{I,E} + \Delta V_{I,E}] \,\bar \delta(V_{I,E} + \Delta
V_{I,E}), \ee where $\Delta V_{I,E}$ is a small change in the size of the
radius.  Working to first order in $\Delta V_{I, E}$, this gives \be \delta =
\bar \delta + V_{I,E} \frac{d \bar \delta}{d V_{I,E}}.
\label{eq:deltaofVE}
\ee 
We can therefore rewrite equation~(\ref{eq:meanscreen_infall}) using the
Lagrangian volume as 
\ba \frac{dV_I}{dt} &=&
\frac{1}{\left[(1-f_{\rm coll})\chi_{\rm eff}+[1+A({\rm He})]f_{\rm coll,MH}\bar\xi \right]}\nonumber \\
&\times&\left[\frac{\dot{N}_\gamma}{n_H^0} - \alpha_B \, C \, (1+z)^3 \, n_H^0
  \chi_{\rm eff}\int_{V_0}^{V_I} dV_I' \delta(V_I')^{-1}
  \delta(V_I')^2\right].
\label{eq:meanscreen_infall2}
\ea 
The relevant overdensity that appears in
equation~(\ref{eq:meanscreen_infall2}) is 
\be 
\delta_{\rm clump} (V_I) \equiv
\frac{1}{V_I-V_0} \int_{V_0}^{V_I} dV_I' \delta(V_I').
\label{eq:deltaclump}
\ee 
In Figure~\ref{overdensities}, we show $\bar \delta$, $\delta$, and
$\delta_{\rm clump}$ around peaks of mass $10^{8} \msun$ and
$10^{11} \msun$.  Note
that $\delta_{\rm clump}$ exceeds $\bar \delta$ over a range of radii, because
$\delta_{\rm clump}$ is an average in Lagrangian coordinates and $\bar \delta$
is an average in Eulerian coordinates.  Finally, the flux in
equation~(\ref{flux}) is corrected by a similar factor of $\delta_{\rm
  clump}$, yielding 
\be 
F=\frac{{\dot{N}_\gamma} - \alpha_B \, C \, (1+z)^3 \,
  (n_H^0)^2 {\chi_{\rm eff}}\, (V_I-V_0) \delta_{\rm clump}(V_I)}{4\pi
  a^2r_{I,E}^2}.
\label{flux_infall}
\ee

The average  number of ionizing photons absorbed per 
minihalo atom, $\bar \xi$, in the process of evaporating all the minihalos
at the current location  of the I-front, $r_I(t),$ must now be specified.
\citet{ISR04} have shown that the number of photons per atom
absorbed by a minihalo of mass $M_7$ (in units of $10^7 M_\odot$), 
overtaken by an intergalactic I-front at 
a redshift of $z$ which is driven by an external source of flux
$F_0$, the flux in units of that from a source emitting 
$N_{\rm ph}=10^{56}\rm s^{-1}$ ionizing photons per second at a proper 
distance $d$ of 1 Mpc, i.e.  
\be
F_0             \equiv \frac{N_{\rm ph,56}}{d^2_{\rm Mpc}}
        =\frac{F}{8.356\times10^5\rm s^{-1}cm^{-2}},        
\label{eq:Fclump}
\ee 
during its photoevaporation is given by 
\be
\xi(M,z,F_0)\equiv 1+\phi_1(M)\phi_2(z)\phi_3(F_0) 
\ee 
where $\phi_1(M) \equiv
A(M_7^{B+C\log_{10} M_7})$, $\phi_2(z)\equiv \left[G + H\left({1+z}\right)/10
\right]$ and $\phi_3(F_0) \equiv F_0^{D + E {\rm log_{10}} F_0}$.  Here the 
factors
$A$-$G$ are dependent on the spectrum of the ionizing sources: $A$ = (4.4,
4.0, 2.4), $B$ =(0.334, 0.364, 0.338), $C$ = (0.023, 0.033, 0.032), $D$
=(0.199, 0.240, 0.219), $E$ = (-0.042,-0.021,-0.036), $G$=(0.,0.,0.1), $H$=(1,
1, 0.9), for the cases in which the ionizing spectrum is taken to be a $5\times
10^4K$ blackbody representing Pop.~II stars , a QSO-like power-law 
spectrum  with slope of $-1.8$, or a $10^5$K blackbody representing 
Pop.~III stars, respectively.

In order to use this result in equation (\ref{eq:meanscreen_infall2}), the
quantity $\xi(M,z,F_0)$ must first be averaged over the mass function
of minihalos at $r_I(t)$ on the undisturbed side of the I-front as the
H~II region evolves with time.  Since $\xi(M,z,F_0)$ depends on
$F_0(t)$, which also depends on $r_I(t)$ according to equation
(\ref{flux_infall}), equations (\ref{eq:meanscreen_infall2}) and
(\ref{flux_infall}) are coupled and must be solved simultaneously.  
We shall consider three analytical approximations for the minihalo mass
function when averaging $\xi(M,z,F_0)$ to obtain  $\bar \xi$.  The first 
two approaches, described in \S~\ref{unbiased_MH}, are based on
the well-known Press-Schechter (PS) approximation for the mass function
averaged over all space at a given redshift \citep{PS74}.  
We shall refer to these, which depend upon $z$ and $F_0$,  but
not upon $r_I(t)$ or the source halo properties, as ``unbiased
minihalo'' averages.  The third approximation, described in 
\S~\ref{biased_MH}, is based upon an
extension of the PS approach which takes account of the spatial
correlation between the minihalos and the central source
halos, as described by \citet{SB02}.  In this last
approximation, which we refer to as the ``biased minihalo'' average,
$\bar \xi$ not only depends upon $z$ and $F_0$, but also on $r_I(t)$
and the source halo mass.

\subsubsection{The Average Photon Consumption Rate for Unbiased Minihalos.}
\label{unbiased_MH}

We begin by defining the average photon consumption rate per minihalo
atom 
\be 
\bar{\xi}_{\rm nb,1}(z,F_0) \equiv \frac{\int_{M_{\rm min}}^{M_{\rm max}} dM \,
  \frac{dn(M,z)}{dM} \, M \, \xi(M,z,F_0)} {\int_{M_{\rm min}}^{M_{\rm max}}
  dM \, \frac{dn(M,z)}{dM} \, M \, },
\label{meanxi}
\ee 
where $\frac{dn(M,z)}{dM}$ is the PS mass function
of halos, 
if we assume that the minihalos at a given redshift $z$ just formed at
that redshift.  If, on the other hand, we assume that minihalos at $z$ had a
distribution of formation redshifts $z_f$, with $z_f \geq z$, then 
\be 
\bar{\xi}_{\rm nb,2}(z,F_0)\equiv
\frac{\int_{M_{\rm min}}^{M_{\rm max}} dM \, \int_z^\infty dz_f \,
  \frac{d^2n(M,z_f)}{dM dz_f} \, \, M \, 
      \xi(M,z_f,F_0)} {\int_{M_{\rm min}}^{M_{\rm
      max}} dM \,\int_z^\infty dz_f \, \frac{d^2n(M,z_f)}{dM dz_f} \, \, M \, }.
\label{eq:meanxi_z}
\ee 
In both these equations, 
the limit $M_{\rm min}$ is the minimum minihalo mass (which we
assume here to be the Jeans mass at that epoch), while $M_{\rm max}=M(T_{\rm
  vir}=10^4 K)$ is the halo mass at that epoch for which $T_{\rm vir}=10^4$ 
K. In equation (\ref{eq:meanxi_z}) we have 
approximately accounted for the distribution of minihalo formation times, by 
taking the derivative of the mass function, which
glosses over the fact that the change in this function at a given mass $M$
includes both a positive contribution from halos whose masses have increased
to $M$ from lower values, as well as a negative contribution from halos whose
masses have increased from $M$ to higher values.  The error introduced by this
approximation is small, however \citep{KS96}, and is justified given the other
uncertainties involved.

\subsubsection{The Average Photon Consumption Rate for Biased Minihalos.}
\label{biased_MH}

To calculate the biased distribution of minihalos about a given source, we
employ an analytical formalism that tracks the correlated formation of
objects.  Our approach, described in detail in \citet{SB02},
extends the standard PS method using a simple approximation to construct the
bivariate mass function of two perturbations of arbitrary mass and collapse
redshift, initially separated by a fixed comoving distance \citep{Poetal98}.  
From this function we can construct the number density of
minihalos of mass $M$ that form at an initial redshift $z$ at a comoving
distance $r$ from the source halo of mass $M_s$ and formation redshift $z_s$:
\be \frac{dn}{d M} (M,z,r|M_s,z_s) = \frac{\frac{d^2 n}{dM dM_s}
  (M,z,M_s,z_s,r)} {\frac{dn}{dM_s} (M_s,z_s)},
\label{eq:biasnum}
\ee where $\frac{dn}{dM_s} (M_s,z_s)$ is the usual PS mass function and
$\frac{d^2 n}{dM dM_s} (M,z,M_s,z_s,r)$ is the bivariate mass function that
gives the product of the differential number densities at two points separated
by an initial comoving distance $r$, at any two masses and redshifts.  Note
that this expression interpolates smoothly between all standard analytical
limits: reducing, for example, to the standard halo bias expression described
by \citet{MW96} in the limit of equal mass halos at the same redshift,
and reproducing the \citet{LC93} progenitor distribution in the limit
of different-mass halos at the same position at different redshifts. 
Further detaled checks of the method against N-body simulations were presented
in \citet{ISMS03} and \citet{ST05}. Note
also that in adopting this definition we are effectively working in Lagrangian
space, such that $r$ is the {\em initial} comoving distance between the
perturbations. As a shorthand we define $\frac{dn_{mh,s}}{d M} (z, r) \equiv
\frac{dn}{d M} (M,z,r|M_s,z_s).$  With this definition the biased values of 
the average photon consumption per minihalo atom corresponding to the two
unbiased cases in equations~(\ref{meanxi}) and (\ref{eq:meanxi_z}),
respectively, are
\be 
\bar \xi_{b,1}(z,F_0,r_I,M_s) = 
\frac{\int_{M_{\rm min}}^{M_{\rm max}} dM 
  \, \frac{dn_{mh,s}}{dM} (z,r_I) \, M \, \xi(M,z,F_0)} {\int_{M_{\rm
      min}}^{M_{\rm max}} dM \, \, \frac{dn_{mh,s}}{dM} (z,r_I) \, M \,}
\label{eq:biasxi}
\ee 
and
\be 
\bar \xi_{b,2}(z,F_0,r_I,M_s) = 
\frac{\int_{M_{\rm min}}^{M_{\rm max}} dM \int_z^\infty
  dz_f \, \frac{d^2n_{mh,s}}{dM dz_f} (z_f,r_I) \, M \, \xi(M,z_f,F_0)} {\int_{M_{\rm
      min}}^{M_{\rm max}} dM \, \int_z^\infty
  dz_f \, \frac{d^2n_{mh,s}}{dM dz_f} (z_f,r_I) \, M \,,}
\label{eq:biasxi_z}
\ee 
with $r_I$ the Lagrangian radius of the I-front.

\begin{figure*}
  \includegraphics[width=3.2in]{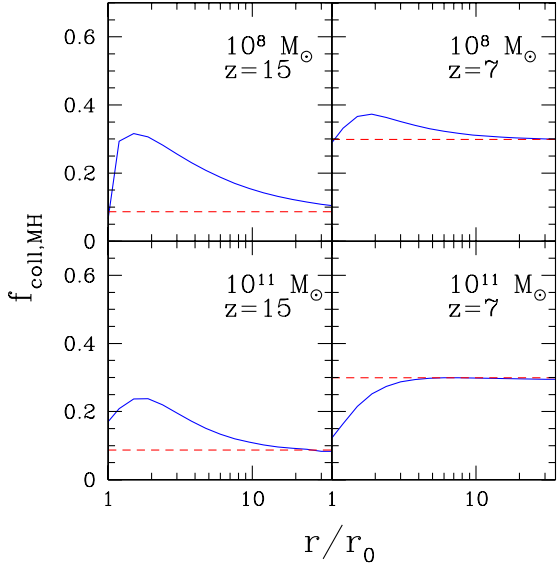}
\caption{Biased collapsed fraction of baryons in minihalos, 
$f_{\rm coll,MH},$ 
as a function of the Lagrangian distance from the source halo
(in units of the source halo Lagrangian radius) for
sources of masses $10^8M_\odot$ and $10^{11}M_\odot$ and redshifts
$z=15$ and $z=7$, as indicated (solid). For reference we also show the 
unbiased collapsed  fraction of baryons in minihalos at the 
corresponding redshifts (dashed).(See the electronic 
edition of the Journal for the color version of this figure.)}
\label{screen_fig}
\end{figure*}

We can anticipate how important this bias effect is likely to be in
determining the average minihalo consumption rate by considering 
its impact on the minihalo collapsed fraction in the vicinity of
a given source halo.  In Figure \ref{screen_fig}, we plot the
average (unbiased) minihalo collapsed fraction versus the biased value,
which varies with distance from the source halo, for two halo masses
$10^8 \msun$ and $10^{11} \msun,$ at redshifts $z=7$ and $z=15.$  
These halo masses and redshifts illustrate the range of behavior
expected during reionization.   In terms of the Gaussian-random-noise
initial conditions for our $\Lambda$CDM cosmological model, 
$10^8 \msun$ halos correspond 
to fluctuations that are 3.2 $\sigma$ (1.6 $\sigma$) at $z=15$ (7),
respectively while $10^{11} \msun$ halos are 6.4 
$\sigma$ (3.2 $\sigma$) at these same redshifts.

The distance between sources and minihalos
is measured in terms of the comoving Lagrangian
(i.e. unperturbed) radius of a given mass shell surrounding the
central source, in units of $r_0$, the Lagrangian radius of the source. 
According to this figure, the bias can be significant for
minihalos located in the range $ 1 \lesssim r/r_0 \lesssim 10.$  For
typical source halos, in fact, the biased collapsed fraction in
the neighborhood of the source hardly declines with increasing redshift,
in contrast with the unbiased collapsed fraction which declines by
a factor of more than $3$ between $z=7$ and $z=15$.

\subsection{The Ionizing Photon Luminosity of the Central Source}

The total ionizing photon output of a source, $N_\gamma$, and its time
evolution depend on the mass of the host halo $M_s$, photon production per
stellar baryon $N_i$, star-formation efficiency $f_*$ 
and ionizing photon escape
fraction $f_{\rm esc}$.  We then define the total ionizing photon output per
source atom that escapes the source halo as: \be f_\gamma=f_*f_{\rm esc}N_i,
\ee thus the source emits a total of $N_\gamma=f_\gamma M\Omega_b/(\mu m_p)$
during its lifetime, where $\mu m_p$ is the mean mass per atom.

The star-formation efficiency $f_*$ and ionizing photon escape fraction
$f_{\rm esc}$  are highly uncertain in general, and even more so for
the high-redshift galaxies responsible for reionization. Their estimated
values vary by several orders of magnitude between different observational and
theoretical estimates \citep{LFHL95,RS00,Hetal01,Setal01,TM01}.  For
simplicity, and since the principle aim of this investigation is to show the
effect of small-scale structure rather than model the sources in detail, we
assume that each source produces a fixed number $f_\gamma$ of ionizing photons
which escape from the source galaxy per atom in the source during the
source's lifetime. The ionizing photon production per atom for Pop.~II
low-metalicity stars with a Salpeter IMF is $N_i=3000-10000$
\citep{Letal99}.  Zero-metalicity, massive Pop.~III stars, on the other hand,
are estimated to produce values of $N_i$ that rise sharply with mass from
25,000 to 80,000 as stellar mass increases from 10 $\msun$ to 50 $\msun$,
then gradually reach a peak of 90,000 at 120 $\msun$, and finally
decline slowly to 80,000 by 500 $\msun$ \citep{S02,TVS04}. 
While individual Pop.~III stars have higher values 
of $f_\gamma$ than  Pop.~II stars of the same mass, a non-trivial part of the
increase from Pop.~II to Pop.~III quoted above reflects the fact that the
assumed Pop.~II IMF has many low-mass stars, which are inefficient ionizing sources, 
while the Pop.~III IMF is often hypothesized to contain only massive stars.

Assuming, conservatively, that $N_i \geq 4000$ 
for Pop.~II stars and $N_i \geq 25,000$
for Pop.~III stars, and taking moderate fiducial values for the photon escape
fraction and star formation efficiency of $f_{\rm esc}=0.1$ and $f_*=0.1,$
yields $f_\gamma \geq (40,250) (f_{\rm esc}/0.1) (f_*/0.1)$
for (Pop.~II, Pop.~III), respectively. 
We shall further assume that the time-dependence of this ionizing photon 
output is characteristic of a starburst with a photon luminosity
\be
  \dot{N}_\gamma=f_\gamma\frac{\alpha-1}{\alpha}\frac{M\Omega_b}{\mu m_p
    t_s}\times \left\{ \begin{array}{ll}
      1 &t\leq t_s\\
      \left(\frac{t}{t_s}\right)^{-\alpha}& t>t_s,
    \end{array}
  \right.   \ee  where $\alpha=4.5$, i.e. we assume that the source is
steady for a time $t_s$, after which the photon flux decreases as
power of time \citep{HH03}.  Here $t_s$ is the characteristic time for a
source to fade, essentially the typical source lifetime. For
individual massive stars $t_s\approx 3$~Myr, but starbursts could in
principle last significantly longer, thus we consider both the cases
$t_s=3$~Myr and $t_s=100$~Myr. We assume that the He correction
$\chi_{\rm eff}$ is 1.08 for the softer Pop.~II spectrum and 1.16 for
the harder Pop.~III and QSO spectra.

\subsection{Results for Individual H~II Regions}
\label{results}

\begin{figure*}
  \includegraphics[width=3.2in]{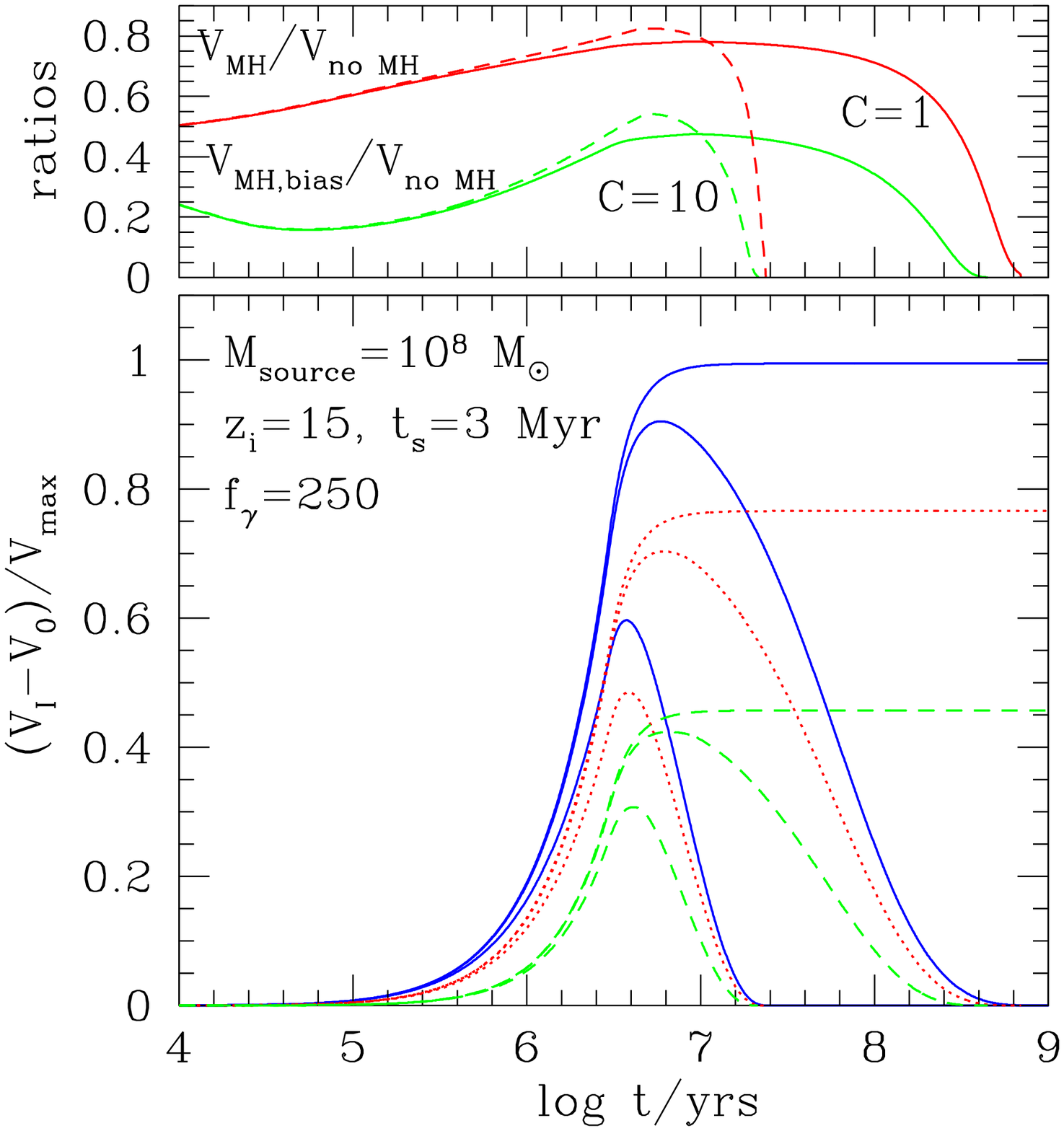}
  \includegraphics[width=3.2in]{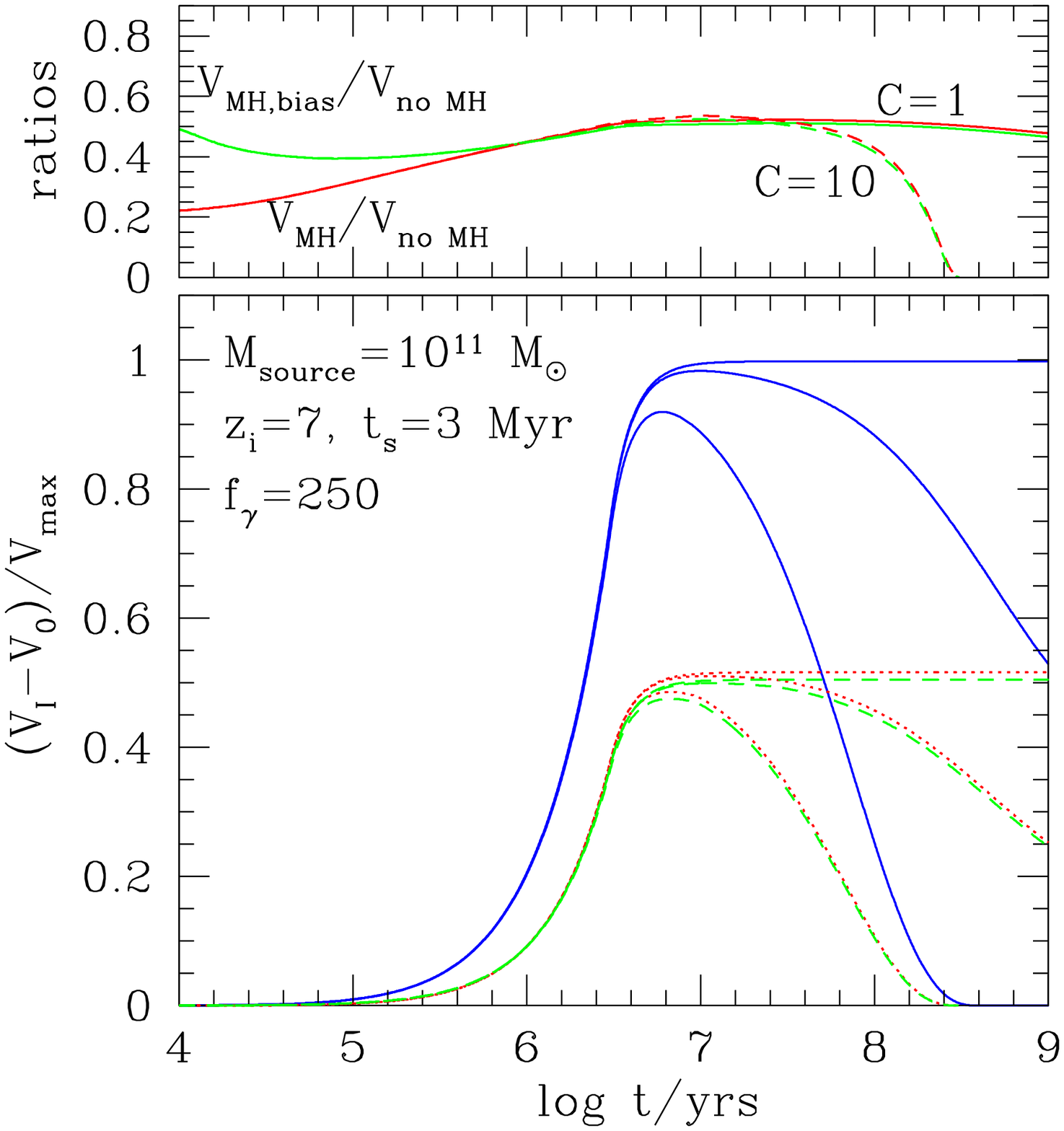}
\caption{{\em Bottom:} 
  The evolution of the Lagrangian volume of the H~II region about a single
  source  of (left) mass $10^{8}M_\odot$ that turns on at $z_i=15$,
  or (right) mass  $10^{11}M_\odot$ that turns on at $z=7$.  Both sources
  have
  Pop.~II  stellar spectra and lifetimes of $t_s=3$ Myr, during which they
  produce a total of $f_\gamma=250$ photons/atom.  Shown
  are the cases of no minihalos (solid), unbiased minihalos (dotted),
  and biased minihalos (dashed) for IGM clumping factors (top to bottom
  in each case) $C=0$ (i.e. no recombinations in IGM gas), 1 (mean
  IGM), and 10 (clumped IGM).  $V_{\rm max}$ is the maximum ionized
  volume reached during the lifetime of the source in the $C=0$ case
  with no minihalos, as defined in the text. {\em Top:} Ratios of the
  ionized volumes with unbiased and biased minihalos to the no
  minihalo case, as labeled, for $C=1$ (solid) and 10 (dashed).(See the electronic 
edition of the Journal for the color version of this figure.)}
\label{vol}
\end{figure*}

\begin{figure*}
  \includegraphics[width=3.2in]{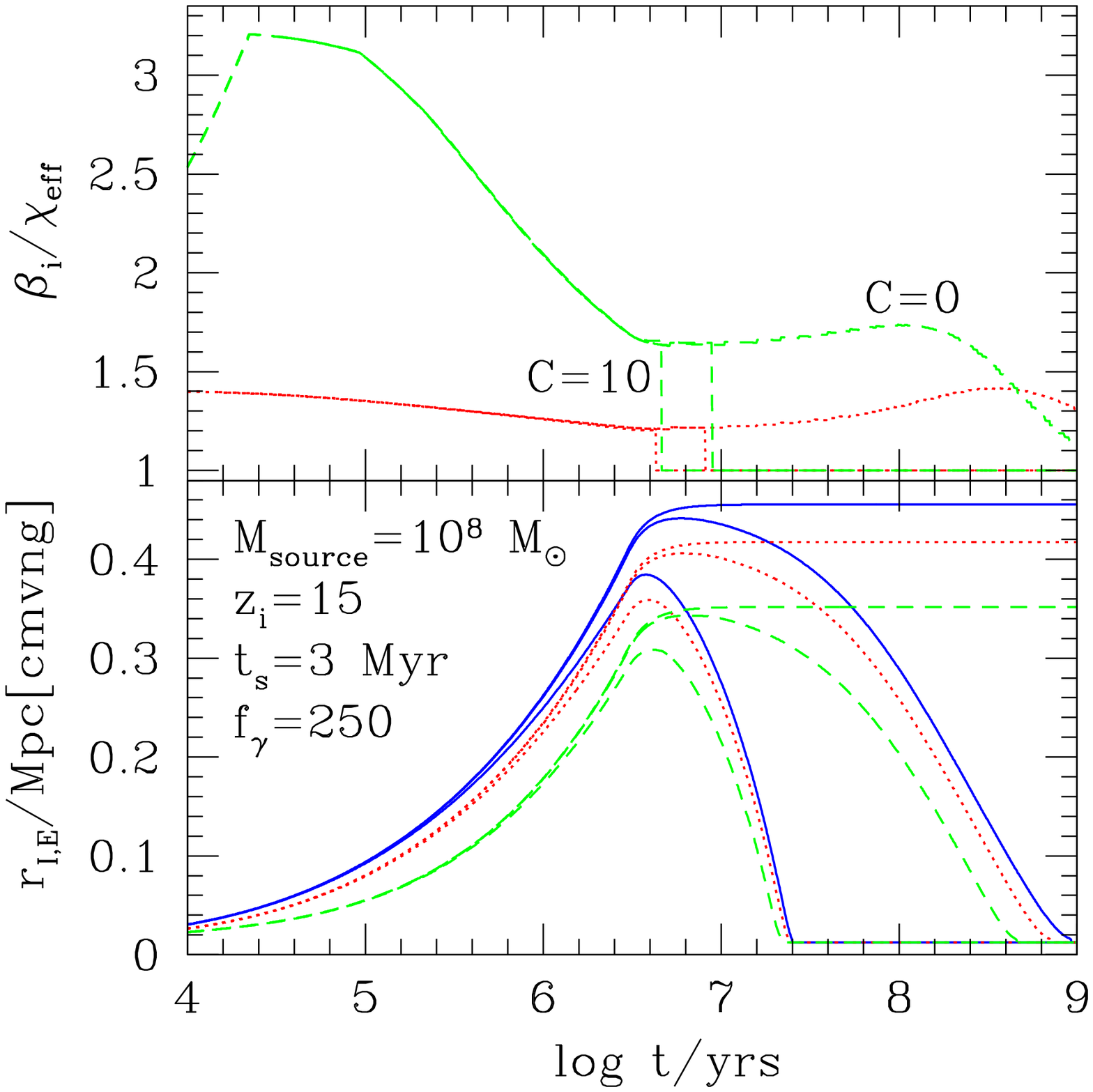}
  \includegraphics[width=3.2in]{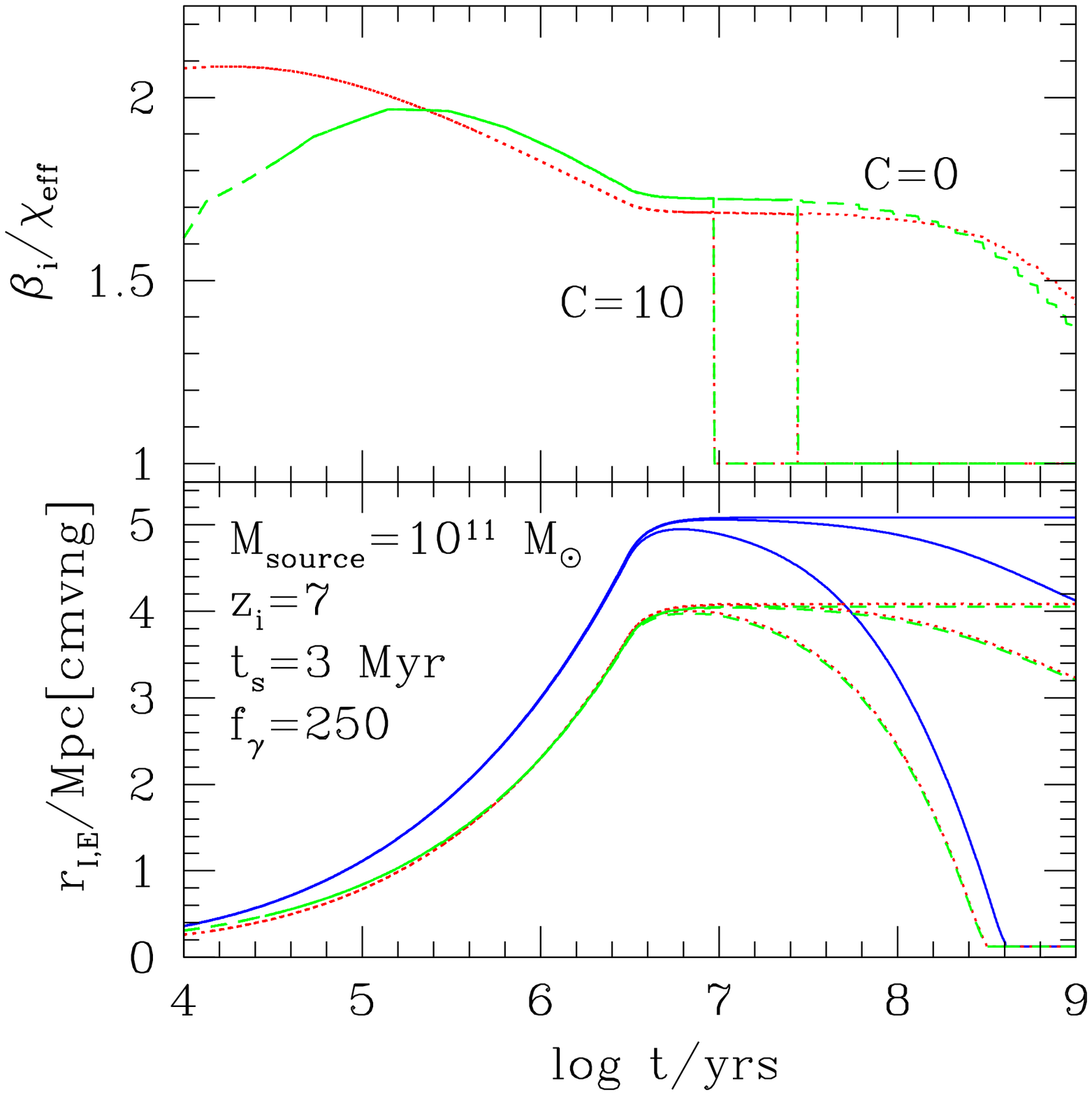}
\caption{Evolution of individual H II regions with the same source parameters
  as in  Figure~\ref{vol}. {\em Top:} The correction factor 
$\beta_i/\chi_{\rm eff}$ due to minihalos for the number 
of ionized   photons consumed per atom that crosses the I-front, 
for biased (dashed) and
  unbiased minihalos (dotted) for $C=0,1$ and 10. {\em Bottom:}
  Comoving radius of the H II region for no minihalos (solid), unbiased
  (dotted) and biased minihalos (dashed) for $C=0,1$ and 10. (See the electronic 
edition of the Journal for the color version of this figure.)}
\label{xi}
\end{figure*}

\begin{figure*}
  \includegraphics[width=3.2in]{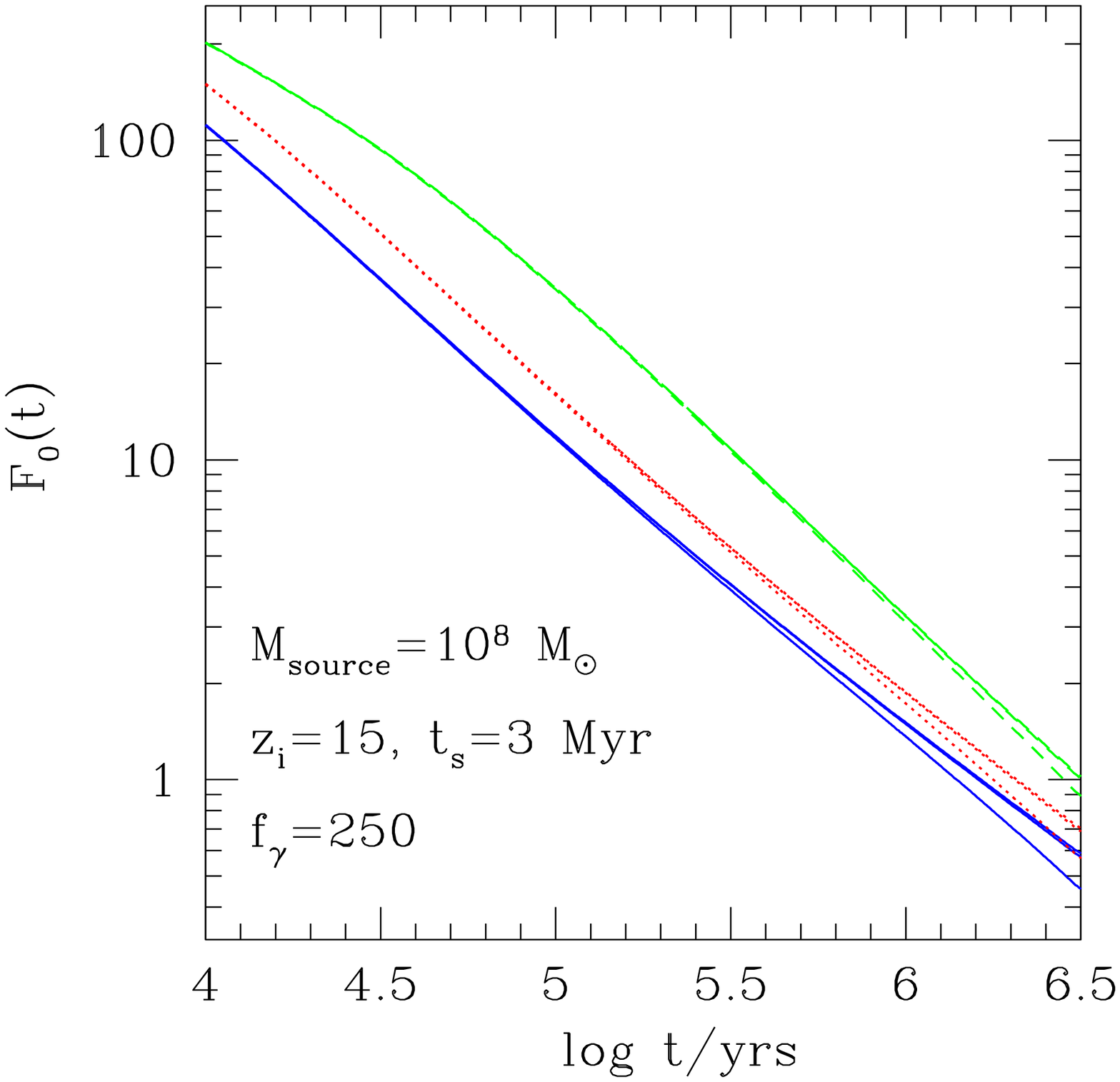}
  \includegraphics[width=3.2in]{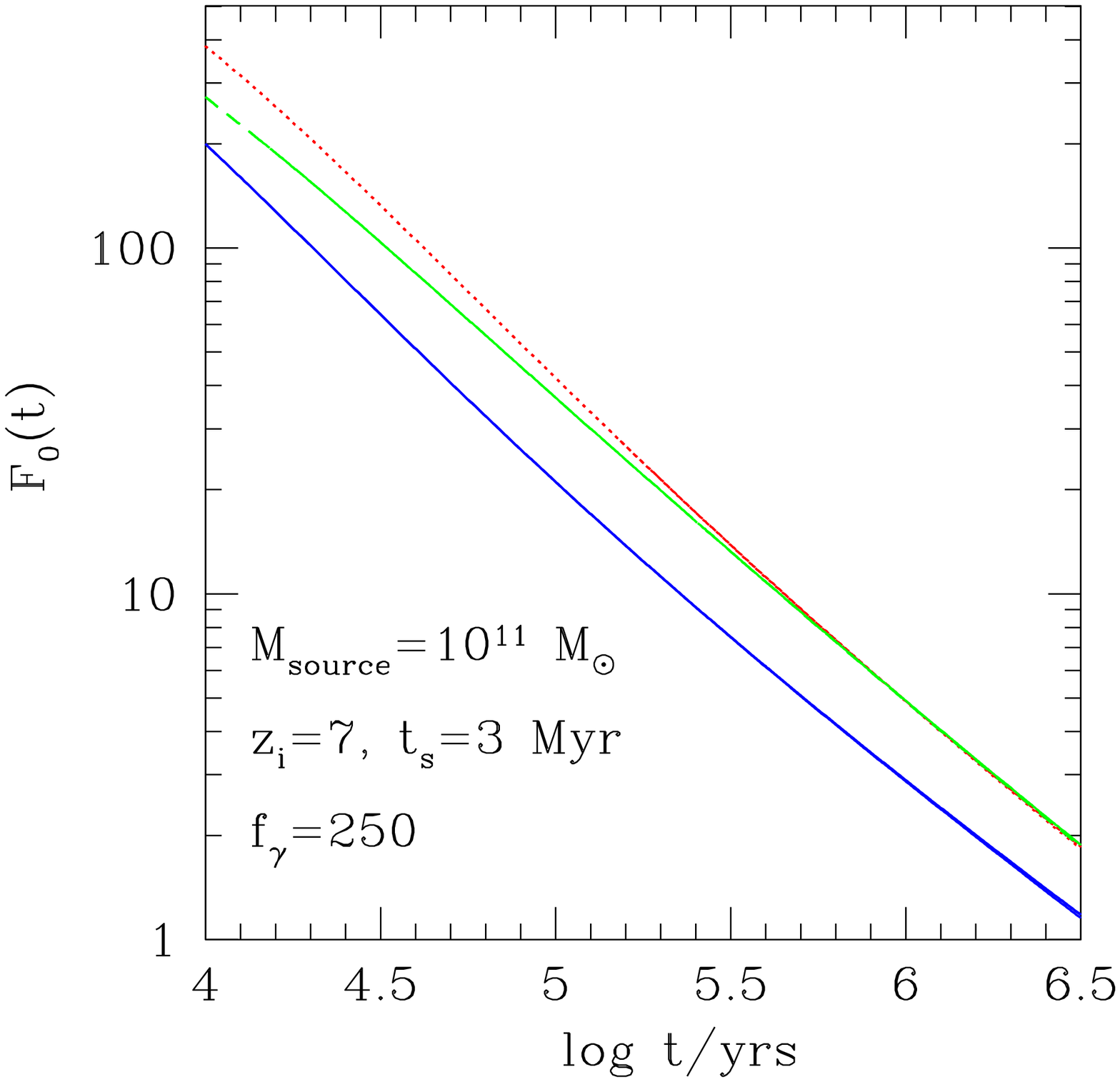}
\caption{Evolution of the dimensionless ionizing photon flux $F_0(t)$
  at the  current position of the I-front. Same notation as in
  Fig.~\ref{vol}. Bottom set of (initially-overlapping) curves are for 
$C=0$, middle set - for $C=1$ and top set - for $C=10$, respectively.(See the electronic 
edition of the Journal for the color version of this figure.)}
\label{flux_fig}
\end{figure*}

\begin{figure*}
  \includegraphics[width=3.5in]{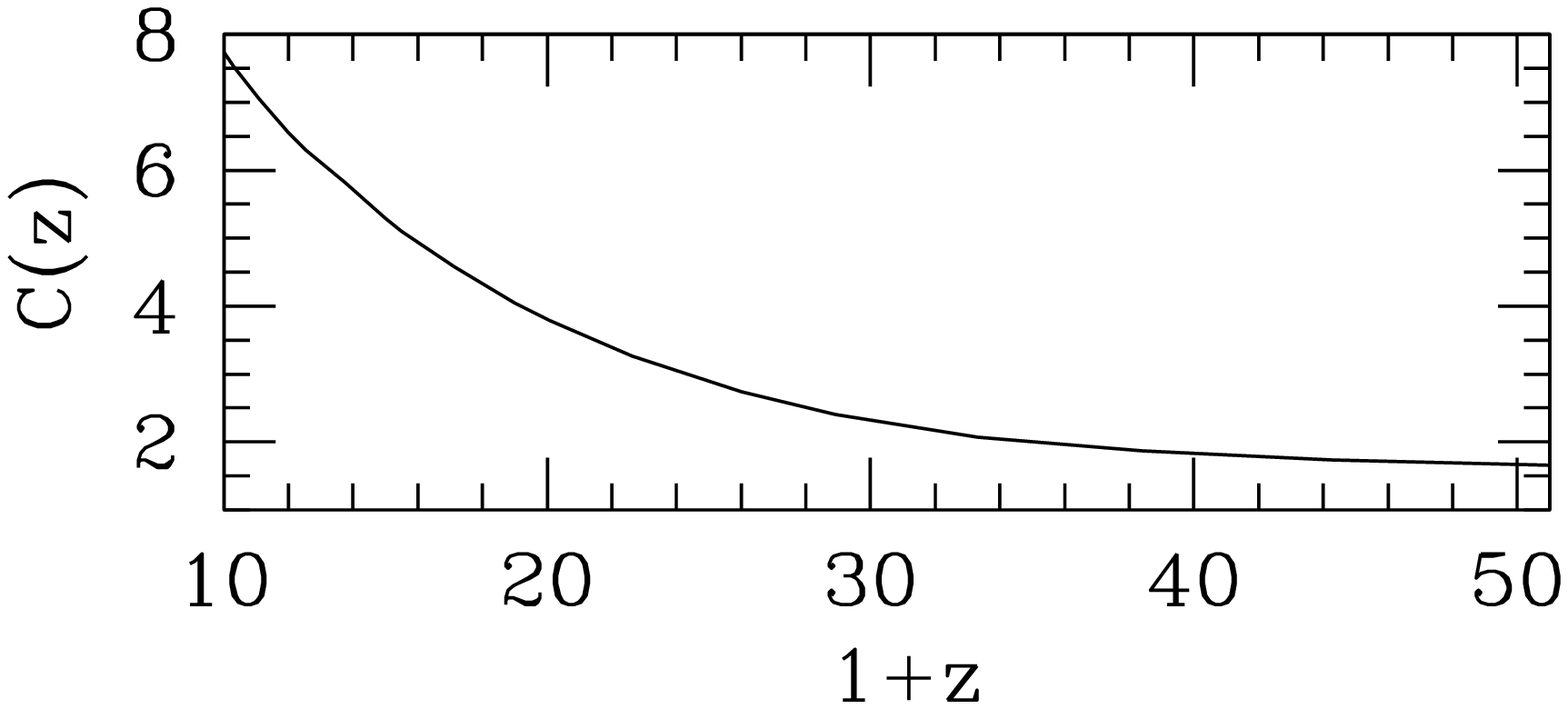}
\caption{Evolution of the IGM clumping factor in $\Lambda$CDM
  from numerical N-body simulations, for the gas {\em outside} halos.}
\label{clumpfact}
\end{figure*}
We present the results of our numerical solution of the spherical
I-front evolution equations from \S 2.2 for two illustrative cases.
Sample results for two source halos, a $10^8 \msun$ halo that turns
on at $z=15$ and a $10^{11} \msun$ halo that turns on at $z=7$, both
with $f_\gamma = 250,$ $t_s = 3$ Myr, and Pop.~II spectra, are given in
Figures~\ref{vol}-\ref{flux_fig}.   In both cases we display results
for a variety of clumping factors ($C=0,1,10$) and successive
approximations of no minihalos, unbiased minihalos [as given by eq.\
(\ref{meanxi})], and biased minihalos [as given by eq.\
(\ref{eq:biasxi})].  In Figure~\ref{vol} we show the evolution of the
ionized volume. All volumes are normalized to $V_{\rm
max}=N_{\gamma}/n^0_H$, i.e. the volume ionized if there were no
minihalos and no recombinations in the gas.

The increase of the clumping factor from $C=1$ to $C=10$ significantly
decreases the maximum ionized volume achieved  by the H~II regions,
with and without minihalos, especially at higher redshift, as shown
for  the $z=15$ case in Figure~\ref{vol}.  For similar reasons, H~II
regions around source halos in overdense areas of the IGM  must also
be smaller than those in the mean IGM.  For $C=0$ and no minihalos,
$V_{\rm max}/V_0 = f_\gamma,$ so $r_{\rm I,max}/r_0 = f_\gamma^{1/3}.$
Since more realistic cases, with $C > 1$ and minihalos, all have $V_I <
V_{\rm max}$, it must be true that $r_I/r_0 < f_\gamma^{1/3},$ in
general.  According to Figure~\ref{overdensities}, the overdensity 
$\delta_{\rm clump} >
1$ outside the source halo for all $r/r_0 \lesssim 20$, while
$\delta_{\rm clump} > 2 $ for all  $r/r_o \lesssim 4$ (6), for source
halos of mass $10^{11}$ $(10^8)$ $\msun$.  As such, the IGM
recombination correction  to the ionized volume at any epoch must be
significantly enhanced by this local overdensity for any $f_\gamma
\lesssim 1000.$  In short, for a  realistic range of $f_\gamma$ 
values, cosmological H~II regions from stellar sources are
generally not larger than the infall regions associated with their
source halos. This is true with and without minihalos.

Next we consider the effect of adding minihalos.  According to 
Figure~\ref{vol}, for the  $10^8 \msun$ source halo, the
unbiased minihalo distribution decreases the ionized volume by $\sim
20\%$ compared to the no minihalos case, while when we account for the
minihalo bias about the source, the ionized volume is decreased by a 
factor of 2 relative to the no minihalos case.
For the $10^{11} \msun$ source halo, the net ionized volume
when minihalos are present (biased or not) is about 65\% of the 
volume in the case without minihalos. 

In the top panel of Figure~\ref{xi}, we plot the factor 
$\beta_i/\chi_{\rm eff}$ by which minihalo evaporation boosts the number
of ionizing photons consumed at the I-front per atom that crosses
the front (in the IGM and in minihalos combined). In the bottom panel of 
this figure, we plot the comoving (Eulerian) radius of the corresponding
H~II regions.  For the $10^8 \msun$ source halo at $z_i = 15,$ 
ignoring the minihalo bias (bottom lines) seriously
underestimates the photon consumption by a factor of $\gtrsim 2$ as
compared to the biased minihalos (top lines).  Furthermore, the
overall effect of adding minihalos is to increase the photon
consumption by $\sim100\%$. For the $10^{11} \msun$ halo at $z_i = 7$,
there is a similar increase in the photon consumption, but little
difference  between the biased and unbiased results.
  
Finally we plot the dimensionless flux $F_0$ at the current position 
of the I-front, in Figure~\ref{flux_fig}.  This is important as a 
check of our assumptions, which incorporate simulation results for minihalo 
evaporation for a range of fluxes, $10^{-2} \leq F_0 \leq 10^3$.
The value of $F_0$  has a significant impact on the
ionizing photon consumption of minihalos, which is higher for 
higher values of $F_0.$ In both cases, the
flux starts fairly high ($F_0 \gtrsim 100$), but drops to
$F_0\sim1$ by the time the source starts to fade.  
At first this drop is due mainly to geometric dilution,
but later, recombinations in the ionized volume accelerate this
decrease according to equation~(\ref{flux_infall}).

\section{Towards a More Global Picture}
\label{global_sect}

To construct models of the global reionization process, we first
calculate the number density of source halos at each redshift based on
the PS formalism. We then 
calculate the evolution of the H~II region created by each source halo,
as discussed in section \S2.2 and \S2.3.  Finally, we 
add the volumes of all these H~II regions.
This gives the total ionized
mass fraction at each redshift, according to 
\be 
f_{I,M}=\int_{M_{\rm
s,min}}^\infty dM\int_z^\infty dz'\frac{d^2n_s(M,z')}{dMdz'}
V_I(z,z',M,\dot{N}_\gamma),
\label{source_int} 
\ee 
where $\frac{d^2n_s(M,z')}{dMdz'}$ is the PS distribution of the source
halos, and we assume here that source halos have masses
$M_s \geq M_{\rm s,min}= M(10^4\rm K),$ the mass
of halos with $T_{\rm vir} = 10^4$ K.  The universe is fully ionized
when the H~II regions overlap, which corresponds to $f_{I,M}=1$.   

\subsection{Time-Dependent Clumping Factor of IGM Outside Halos}

As in the individual-source results above, we calculate the evolution 
of the ionized mass fraction for the cases of no minihalos, unbiased
minihalos, and biased minihalos, and for clumping factors of $C=0, 1$
and $10$. In addition, we consider the more realistic case of a clumping
factor that evolves in time as more and more structure forms, which
was obtained from numerical N-body simulations by the  
particle-particle/particle mesh (P$^3$M) method, with a computational
box size of 1 comoving Mpc with $128^3$ particles and $256^3$ cells,
corresponding to a particle mass of $2\times10^4M_\odot$ (see Shapiro
2001 and Iliev et al.\ 2003 for details on the simulations). 
This box size was chosen so as to resolve the scales which contribute 
most of the clumping - on smaller scales the gas would be Jeans smoothed, 
while on larger scales the density fluctuations are still linear and do not 
contribute much to the overall clumping factor. The result for the IGM 
clumping factor is plotted in Figure~\ref{clumpfact}. This clumping factor
excludes the matter in collapsed halos, since as we discussed above,
these are self-shielded and we treat them separately.
The evolution of this IGM clumping factor with redshift is well-fit by
\be C(z)=17.6 e^{-0.10z+0.0011\,z^2}.
\label{clumpfact_fit}
\ee

\subsection{The Global Consumption of Ionizing Photons 
  During the Epoch of Reionization}

The reionization of the universe was complete when the volume of all
the H~II regions at some epoch equaled the total volume.  We call
that the epoch of overlap, at redshift $z_{\rm ov}.$ Generalizing
\citet{SG87} we define a useful dimensionless ratio of total number of
ionizing photons emitted per hydrogen atom in the universe until
overlap at $z=z_{\rm ov}$ given by,
\be 
\zeta_{\rm ov}=\frac{f_\gamma\Omega_b}{n_H^0\mu
m_p t_s} \int_\infty^{z_{\rm ov}}dz\int_{M_{\rm s,min}}^\infty M_s dM_s
\int_{t(z)}^{t(z_{\rm ov})}dt'\frac{d^2n_x^0(M_s,z')}{dM_sdz'}\frac{dz'}{dt'},
\label{zeta_def}
\ee where $d^2n_x^0(M,z)/{dMdz}$ is the comoving differential number density of
the source halos of mass $M$ formed at redshift $z$. 

\subsection{Electron Scattering Optical Depth Through the Reionized Universe}

For any given reionization history, the mean optical depth along a
line-of-sight between an observer at $z=0$ and a redshift $z$ due to Thomson
scattering by free electrons in the post-recombination universe is given by  
\be
\tau_{\rm es}(z)=c\sigma_T \int^0_{z} dz' n_e(z') \frac{dt}{dz'},
\label{tau_general}
\ee 
where $\sigma_T=6.65\times10^{-25}\rm cm^{2}$ is the Thomson scattering
cross-section, $c$ is the speed of light, and $n_e(z)$ is the mean number 
density of free electrons at redshift $z$, given by 
\be 
n_e(z)=n_{\rm e,0}(1+z)^3f_{I,M}(z),
\label{n_ez}
\ee 
where $f_{I,M}(z)$ is the ionized fraction of the IGM at 
redshift $z$, and $n_{\rm e,0} = n_H^0 \chi_{\rm eff}$. For comparison with
the value of $\tau_{\rm es}$ between us and the surface of last scattering
inferred from measurements of the polarization of the CMB, we should
evaluate $\tau_{\rm es}(z)$  at $z=z_{\rm rec}$, the redshift of recombination.
For this paper we assume that He is ionized to He~II at the same time as H
is ionized. We make the reasonable approximation that $p=2$ (1) in 
$\chi_{\rm eff}$ for
$z\leq3.5$ ($z>3.5$). In fact, the reionization of helium to He~III at
$z\sim3-4$, inferred from  observations of quasar absorption spectra,
has only a small effect on the total electron-scattering
optical depth and can be ignored for most purposes.
In the case of $f_{I,M}=\rm const$ between us and redshift $z$ (e.g. for an
IGM fully ionized and free of minihalos since $z_{\rm ov}$, $f_{I,M}=1$ for
$z\leq z_{\rm ov}$), the integral in equation~(\ref{tau_general}) has a closed
analytical form  
\be 
\tau_{\rm es}(z)=\frac{2c\sigma_T\Omega_b\rho_{\rm crit,0}}{3H_0\mu_H m_p\Omega_0}
f_{I,M} \chi_{\rm eff} \left\{[\Omega_0(1+z)^3+
\Omega_\Lambda]^{1/2}-1\right\},
\label{tau_ionized}
\ee 
where $\mu_H = 1 + 4 A({\rm He}) = 1.32,$ the mean molecular weight per
H atom. For $f_{\rm I,M}=1(0)$ for $z\leq z_{\rm ov}$  ($z> z_{\rm ov}$),
$\tau_{\rm es}\geq 0.04$ for $z_{\rm ov}\geq6$.  

\section{Global Reionization Models: Results and Conclusions}
\label{summary_sect}
\subsection{Fiducial Model}
\begin{figure*}
  \includegraphics[width=3.2in]{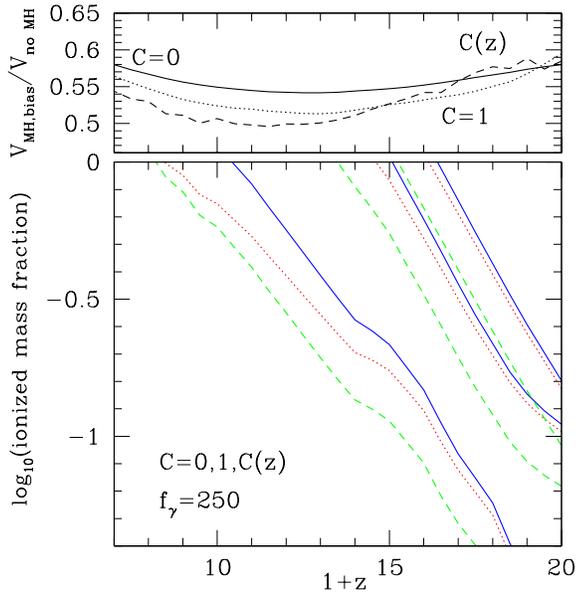}
\caption{Global reionization. Sources with $f_\gamma=250$ and lifetime
  of $t_s=3$ Myr (Case A) are assumed. (bottom panel) Decimal
  logarithm of ionized mass (or Lagrangian volume) fractions (i.e. 0
  corresponds to overlap) for the cases of no minihalos (solid),
  unbiased minihalos (dotted), and biased minihalos (dashed) for IGM
  clumping factors (top to bottom in each case) $C=0$, 1, and C(z)
  (clumped IGM). (upper panel) Ratios of the ionized volume fractions
  in the presence of {\em biased} minihalos and with no minihalos for $C=0$, 1
  and z-dependent.(See the electronic 
edition of the Journal for the color version of this figure.)}
\label{reion}
\end{figure*}

We first consider a fiducial model, Case A, that assumes that the sources are
Pop.~II starbursts, with $f_\gamma=250$ and a lifetime of $t_s=3$ Myr.
The evolution of the ionized mass fractions for Case A, with and without
minihalos and IGM clumping,
is shown in Figure \ref{reion}.  At $z \gtrsim 15$ there are very few
minihalos.  Thus in the $C=0$ case, the no-minihalos model and the
model with unbiased minihalos are almost equivalent.  The few
minihalos that do exist at such early times, however, are highly
biased towards the sources, giving rise to a substantial difference
between the biased and unbiased models.  Similar trends are visible in
the $C=1$ case with minihalos having an appreciable impact only in the
biased case.  On the other hand, in the $C(z)$ model, recombinations
in non-virialized structures slow reionization to the point where a
significant average minihalo collapse fraction builds up, giving rise 
to a shift (i.e. delay) of $\Delta z \sim 2 $ between the no minihalos 
and unbiased minihalo models.  
In this case, the minihalo correction increases with time relative to
the case without minihalos, as expected since the unbiased $f_{\rm coll,MH}$
grows with time.  This has the effect of slowing the rate of increase
of the ionized volume more and more over time.  This trend would help
reconcile an early onset of reionization with a late epoch of overlap.  
When minihalo bias is taken into account, however, the minihalo correction
is just as large at early times as at late times.  At late times
the biased and unbiased minihalo curves almost converge,
since the clustering of small objects is very
weak at these times and bias has a minimal effect.

\begin{figure*}
  \includegraphics[width=3.2in]{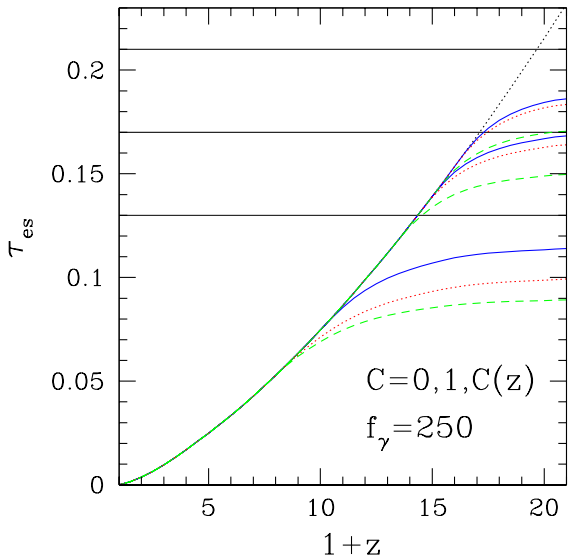}
\caption{Global reionization. Integrated optical depth due to electron
  scattering $\tau_{\rm es}$ vs. redshift. Same notation as in
  Fig.~\ref{reion}. Top (dotted) curve shows the optical depth produced
  assuming complete ionization out to the corresponding redshift. Horizontal
  lines indicate the best-fit and $1-\sigma$ uncertainties of the first-year
  WMAP result, $\tau_{\rm es}=0.17\pm0.04$. (See the electronic 
edition of the Journal for the color version of this figure.)}
\label{tau}
\end{figure*}

The results for $z_{\rm ov}$, $\zeta_{\rm ov}$ and the net $\tau_{\rm es}$,
$\tau_{\rm es}(z_{\rm rec})$, for our 
fiducial model, Case A,  are shown in Table~1 and in Figure~\ref{tau}
for $\tau_{\rm es}(z)$. When there are no minihalos and no recombinations 
in the IGM, $z_{\rm ov} =15.2$ and $\tau_{\rm es} =0.186,$ which requires
just one photon per atom (i.e. $\zeta_{\rm ov} = 1$).
The presence of minihalos by themselves delays overlap
until $z_{\rm ov}=14$, while increasing the global photon consumption
by a factor of $\sim2$ and decreasing the optical depth to 0.169. The
IGM clumping by itself delays overlap until $z_{\rm ov}=8.1 (9.5)$ for
$C=10 [C(z)]$, increasing the global photon consumption to $23 (14)$
and decreasing $\tau_{\rm es}$ to $0.090 (0.114)$. When the effects of both
minihalos and redshift-dependent IGM clumping are included, overlap is
further delayed until $z_{\rm ov}=7.2$, in rough consistency with the results
from the Gunn-Peterson trough observations, while the
electron-scattering optical depth decreases to $0.089$, somewhat below
the $1-\sigma$ WMAP limit, and the global ionizing photon consumption
rises to $\zeta_{\rm ov} =33$.

\subsection{Impact of Model Uncertainties and Discussion}

In order to estimate the effects of small-scale structure under
different assumptions, we consider several cases in addition to our
fiducial model  (Case A).  Case B is the same as our fiducial model,
but assumes that the sources are longer-lived, with $t_s=100$ Myr.
Case C is also  the same as our fiducial model, but assumes that the
formation times of the minihalos are distributed in redshift, by
replacing equation  (\ref{meanxi}) with equation  (\ref{eq:meanxi_z})
for unbiased minihalos and equation~(\ref{eq:biasxi}) with equation 
(\ref{eq:biasxi_z}) for biased minihalos as discussed in sections 
\S~\ref{unbiased_MH} and \S~\ref{biased_MH}.  Cases D, E and F are like
Case A, except with $f_\gamma=40$ (Case D), 150 (Case E) and 500 (Case F). 
Finally, Case G assumes  Pop.~III sources with a
$t_s =3$ Myr lifetime and $f_\gamma=250$.

\begin{table*} 
\begin{minipage}{\textwidth} 
\caption{Global reionization results for ionizing photon consumption,  
$\zeta_{\rm ov}$ overlap epoch, $z_{\rm ov}$ \tablenotemark{a}, and  
electron scattering optical depth, $\tau$.} 
\label{zeta_table} 
{\footnotesize 
\begin{tabular}{@{}l|l|c|c|c|c|c|c|c|c|c|c|c|c} 
\hline
No MH       &                 &          &     $C=0$    &     & &     $C=1$    &        &            & $C=10$        &           &         & $C=C(z)$     &        \\\hline
Case&$f_\gamma$&$\zeta_{\rm ov}$&$z_{\rm ov}$&$\tau_{\rm es}$&$\zeta_{\rm ov}$&$z_{\rm ov}$&$\tau$&$\zeta_{\rm ov}$&$z_{\rm ov}$&$\tau_{\rm es}$ &$\zeta_{\rm ov}$&$z_{\rm ov}$&$\tau_{\rm es}$ \\\hline
A           &250              &1&15.4&0.186  &2.1         &14.2          &0.168   &23          &8.1            &0.089      &14       &9.5           &0.114\\
B           &250,long life    &1&13.6&0.157  &1.8       &13            &0.144   &13          &9              &0.096      &9        &9.8           &0.108\\ 
C           &250,z-distr.     &1&15.4&0.186  &2         &14.2          &0.168   &24.5        &8.1            &0.090      &15       &9.4           &0.114\\ 
D          &40               &1&11.3&0.137  &2.7       &9.3           &0.107   &...           &...              &0.045      &...        &...             &0.048\\ 
E           &150              &1&14.4&0.173  &2.3       &12.7          &0.151   &24          &6.5            &0.072      &   21     &    7.0       &0.088 \\ 
F &500 &1 &17   &0.205 &2. &15.5 &0.192 &20 &10.5&0.118&10.0  &12 &0.151\\ 
G &250 &1 &15.4 &0.186 &2.1 &14.2 &0.168 &23 &8.1 &0.089 &14  &9.5 &0.114\\ \hline
MH, no bias &                 &&&&           &             &        &            &               &           &         &              &\\\hline
A           &250              &1.2&15&0.183  &2.5       &13.6          &0.164   &39          &6.7            &0.077      &30       &7.5           &0.099\\
B           &250,long life    &1.2&13.5&0.155&2         &12.5          &0.142   &14          &9              &0.096      &9        &9.8           &0.107\\
C           &250,z-distr.     &1.2&15&0.183  &2.6       &13.7          &0.163   &43          &6.6            &0.076      &34       &7             &0.095\\ 
D           &40               &1.3&11&0.131  &4.4       &8             &0.095   &...           &...              &0.043      & ...       &...             &0.044\\ 
E           &150              &1.2&14&0.169  &3         &12            &0.144&34          &5.5            &0.059      &   34    &     5.5       &0.072\\
F &500 &1.2 &16.5 &0.203 &2.1&15.5 &0.189 &30 &9.5 &0.108 &16.0 &11 &0.141\\ 
G &250 &1.1 &15.2 &0.185 &2.2 &14 &0.166 &31 &7.3 &0.082 &22 &8.3 &0.106\\ \hline
MH, w/bias  &                 &&&&           &              &        &            &              &           &         &              &\\\hline
A           &250              &1.9&14&0.169  &4.3       &12.4          &0.148   &40          &6.6            &0.072      &33       &7.2           &0.089\\
B           &250,long life    &1.6&13&0.148  &2.3       &12.3          &0.138   &14          &9              &0.095      &10       &9.7           &0.106\\
C           &250,z-distr.     &2.1&14&0.169  &4.5       &12.4          &0.148   &44          &6.6            &0.072      &35       &7             &0.087\\
D           &40               &2.0&10&0.117  &5.1       &7.5           &0.085   &...           &...              &0.043      &...        &...             &0.044\\ 
E           &150              &2.0&13&0.154  &4.6       &11.0          &0.129   &34          &5.5            &0.057      &   34    &  5.5         &0.067\\ 
F &500 &1.8 &15.3 &0.192 &3.4&14.5 &0.176 &37 &9   &0.100 &22.0 &10 &0.126\\ 
G &250 &1.5 &14.7 &0.177 &3.2 &13 &0.157 &32 &7.2 &0.079 &24.5 &8 &0.099\\ \hline
\end{tabular}}
\tablenotetext{a}{When no value of $z_{\rm ov}$ is given,
  overlap was not achieved by $z=5.5$.}
\end{minipage}
\end{table*}
The values of $z_{\rm ov}$, $\zeta_{\rm ov}$ and $\tau_{\rm es}$ for each
of these cases are summarized in Table~\ref{zeta_table}. In all cases the
clumping of the IGM gas a 
strong effect on the duration of reionization by significantly increasing
recombination rates and the ionized photon consumption.  Thus increasing 
$C$ from 1 (no clumping) to 10 delays the overlap by $\Delta z\approx 4-6$ and
increases the global ionizing photon consumption $\zeta_{\rm ov}$ by a 
significant factor, between 6 and 17.
The electron-scattering optical depth $\tau_{\rm es}$ decreases
correspondingly, from $\sim0.11-0.17$ (consistent with current WMAP limits)
for $C=1$, to $\sim0.04-0.1$ for $C=10$.  Interestingly, the cases with
evolving clumping factor $C(z)$ yield epochs of overlap similar to those of 
the $C=10$ cases, while
at the same time noticeably increasing $\tau_{\rm es}$, by 
$\Delta\tau_{\rm es}\sim 0.02$ to $\sim0.09-0.11$, reaching better agreement 
with the observational limits from WMAP.

The presence of minihalos additionally delays reionization, by up to $\Delta
z\approx 2.5$. This effect is strongest for short-lived sources, and
almost disappears for very long-lived sources (Case B).  This is because in the
long-lived case, the same total number of photons are produced over a longer
time, and thus the typical flux responsible for evaporating minihalos is 
lower, leading to more efficient photoevaporation in terms of ionizing photon
consumption. Similarly, in the short-lived case, the photon consumption by
minihalos typically raises the global photon consumption per atom until
overlap, $\zeta_{\rm ov},$ by a factor of $\sim2$, while in Case B,
the $\zeta_{\rm ov}$ increases range from 10 to 60\%, depending on the
clumping factor. 

Turning to the question of our assumptions about minihalo formation times, we
find that the differences between cases A and C are negligible in all cases.
Thus we can generally assume that the minihalos just formed at the redshift of
consideration as opposed to formation times that are distributed in redshift.

What is the effect of varying the efficiency for ionizing photon release? 
In Case D, in which reionization is caused by Pop.~II sources
with $f_\gamma = 40$, we find that overlap is achieved before $z=6$ only when
there is no gas clumping ($C=1$), even if no minihalos are present. Hence,
$f_\gamma$ (which depends upon the assumed photon production per stellar
baryon, star-formation efficiency per halo baryon and the ionizing photon
escape fraction) should be significantly larger than this assumed value
(e.g. by replacing the Salpeter IMF by a top-heavy one). For
$f_\gamma=150$ (Case E), the values of $z_{\rm ov}$ and $\tau_{\rm es}$
without minihalos are essentially equivalent to those in our fiducial Case A
when biased minihalos are included, for all values of the clumping factor. The
quantity $\zeta_{\rm ov}$ is smaller in Case E than in Case A, however.

Since Case A has too small a value of $\tau_{\rm es}$ to satisfy the WMAP
constraint, we consider a case with higher efficiency for photon release
$f_\gamma=500$ (Case F), to increase $\tau_{\rm es}$ by making reionization
earlier. With evolving clumping factor $C(z)$ and biased minihalos, Case F
results in $\tau_{\rm es}=0.126$, which is marginally consistent with the
1-$\sigma$ WMAP constraint, but $z_{\rm ov}=10$, too early to match quasar
observations.   

Finally, we illustrate the difference between Pop.~II and Pop.~III ionizing
source spectra in Case G, for which massive Pop.~III stars provide
$f_\gamma=250$. The only difference between Case G and Case A is the source
spectral shape, which results in somewhat more efficient evaporation of the
minihalos. Hence, overlap is slightly earlier, at $z_{\rm ov}=8$, and
$\zeta_{\rm ov}$ is a bit smaller, at 25, while $\tau_{\rm es}$ is a bit
higher, at $\tau_{\rm es}\simeq0.1$, so there is still not a good agreement
with both the quasar and CMB constraints. If we reduce $f_\gamma$ somewhat,
for this Pop.~III case, the overlap redshift would drop, but so would
$\tau_{\rm es}$, so it would not help to match these observations
simultaneously. We conclude from this that this problem is not very sensitive
to our choice of source spectrum. 

In addition, we checked the dependence of our results (for our fiducial Case
A) on our assumption that the halo mass function is well-described by the PS
distribution, by replacing it with the Sheth and Tormen mass function (e.g.
Sheth \& Tormen 2002, hereafter ST) for the case without bias (our bias 
formalism does not apply to ST). The result is that noticeably more 
ionizing sources (i.e. more massive halos) exist at early times, which makes 
the electron scattering optical depth slightly larger and thus a bit closer 
to the one-sigma WMAP range. However, the effect is rather marginal, and, in the 
presence of minihalos and evolving IGM clumping the overlap occurs at almost 
exactly the same time using either PS or ST. In the case with no bias, at least,
 we conclude that using ST instead of PS does not make a significant difference 
for the current reionization constraints. When bias is included we expect that 
the differences will be even smaller, since biased minihalos cluster strongly 
around the sources, thus screening them and consuming more ionizing photons
than in the unbiased case, which should make the higher abundance of sources 
early on less important. It should also be noted that currently ST is not 
well tested at high redshifts, hence it is not clear if it is more precise
than the standard PS model. In fact, the few simulations of high-z small-scale
structure formation that exist seem to agree with PS reasonably well 
\citep{S01,JH01,CDBO04}. 

There are some additional effects which we have neglected here which
we  intend to treat in a future paper. When the luminosity of a
central source  halo inside an H~II region decays, the size of the
H~II region shrinks. Gas  that was once ionized and inside the H~II
region before this, can later find  itself outside the H~II region,
where it no longer receives ionizing radiation from the central
source. In our current treatment, such gas is assumed to  recombine
and must be ionized again before reionization of the universe is
complete. Formation of minihalos in such gas will be suppressed by the
Jeans  mass filtering of the IGM \citep{SGB94,G00b} and,  to a lesser
extent, by the relic entropy of the gas after it cools and  recombines
\citep{OH03}. A similar effect occurs if and when  additional sources
of heating of the IGM are present beyond the ionizing sources we have
considered, such as an X-ray background \citep{MBA03,OH03,GB03}. Our
neglect of these possibilities for the partial suppression of the
minihalo population may overestimate the contribution of minihalos to
ionizing photon consumption before final overlap. However, this effect
is offset by another, perhaps more important, effect which we have
also neglected. We have considered the clustering of minihalos around
source halos, but we should also consider the clustering of source
halos in space and in time. The latter effect makes it possible for
new sources to ionize the pre-existing ionized volume of an H II
region created by one source, preventing the H II regions from ever
shrinking. In fact, if current simulations of reionization are correct
in this regard, then this effect is likely to be important, since
simulations find that ionized zones tend to grow over time and
eventually overlap, rather than shrinking and disappearing and then
being replaced by independent H II regions that form elsewhere
(e.g. Ricotti, Gnedin \& Shull 2002; Ciardi, Ferrara \& White 2003;
Sokasian et al. 2004). A similar point has been made recently by
\citet{FZH04}.  In that case, the suppression of minihalo formation in
pre-ionized zones which recombine and must be ionized again, mentioned
above, is unimportant, since gas which is ionized once, remains
ionized by the continuous exposure to new sources over time. We will
treat these effects in a future paper.

It would appear then that our fiducial model Case A with minihalos is the
best candidate for helping to resolve QSO constraints on $z_{\rm ov}$ with the
WMAP measurements of $\tau_{\rm es}$. For although the model still falls short
of reproducing the observed $\tau_{\rm es}$, minihalos nevertheless form 
{\it on average} in small numbers at early times and in large numbers at late 
times, thereby extending reionization and accumulating scattering optical
depth before overlap. However, this behavior changes dramatically if one
accounts for the strong clustering of high-$z$ minihalos. As shown in
Figure~\ref{screen_fig}, the minihalo collapsed fraction in the neighborhood
of the typical source halos is significantly higher than the
spatially-averaged collapsed fraction of minihalos at the same epoch. Hence, 
although the total number of
minihalos increases with time, the number {\em around sources} remains largely
fixed, pushing the entire process of reionization to later times, without
significantly extending its duration. 

This can be understood from our analytical solution for an H~II region in a
uniform IGM for a steady source, discussed in \S~\ref{single_sect}, if we
assume that $\beta_i$ is independent of time. In that case, the H~II region
volume $V(t)$ in the presence of minihalos equals the volume it would have had
at some time $t$ in the absence of minihalos, at a later time
$t'=t(\beta_i/\chi_{\rm eff})$. This argument shows that the rate of growth of
the ionized volume, $dV_I/dt$, decreases in the presence of minihalos by
the factor $\beta_i/\chi_{\rm eff}$. This tilts the curve of the rise of the 
mean ionized fraction of the IGM with time to a flatter slope, which has the
effect of extending the reionization epoch and delaying overlap.

In order to reconcile the late epoch of overlap implied by observations of
high-z QSO's with the early start of reionization implied by WMAP polarization
results, we would like the rise of the ionized volume to be more rapid at
early times than at late times. This is the effect which {\it unbiased}
minihalos have, according to the curves in Figure~\ref{reion}, because the
average of the factor $\beta_i/\chi_{\rm eff}$ increases with time in that 
case as the average collapsed fraction of minihalos grows. However, when 
minihalo bias is taken into account, the effective average correction factor
$\beta_i/\chi_{\rm eff}$ remains roughly constant in time, thereby reducing
$dV_I/dt$ by the same factor at all times, instead of a factor which increases
over time as required to reconcile the two observations. Thus although 
minihalos have a
large effect on $\zeta_{\rm ov}$, they do not help to reconcile the WMAP and 
high-redshift QSO results. Instead they essentially act as local screens that
reduce the effective number of photons that can impact the IGM at any given 
redshift, given a choice of $f_\gamma.$ A similar slow-down of the growth of 
the ionized volume would  result if we neglected the minihalos but adopted a 
smaller efficiency for the release of ionizing photons by the source halos, 
instead, as seen by comparing Cases A and E.

The evolving clumping factor, on the other hand, does have the desired 
effect of not only extending reionization, but also producing significant 
ionized volume early on, when the IGM clumping is low. This serves to bring 
the electron scattering 
optical depth closer to the WMAP constraint, while also delaying overlap.

We have demonstrated here that small-scale structure, which has generally been
neglected by previous treatments of reionization, can have a substantial
effect on the duration and epoch of completion of reionization. In the future
we can use the approach presented here to explore reionization under more
complex set of assumptions, allowing for the dependence of the ionizing
photon production efficiency and escape fraction on time and halo mass and for
the clustering of source halos, for instance. Our current results suggest, for
example, that an efficiency parameter $f_{\gamma}$ which decreases in time or
with increasing source halo mass, may cause the rise of ionized volume to
decelerate over time, which would help to explain observations of early
reionization onset and late epoch of overlap. 

\acknowledgements We are grateful to Hugo Martel for his help with the 
N-body simulation data and clumping factor determination. We thank 
Andrea Ferrara for helpful comments during the preparation of this
manuscript.  ITI was supported in part by the Research and Training Network
``The Physics of the Intergalactic Medium'' established by the European
Community under the contract HPRN-CT2000-00126.  ES was supported by an NSF
Math and Physical Sciences Distinguished International Postdoctoral Research
(NFS MPS-DRF) fellowship during part of this investigation; his research was
also supported by the National Science Foundation under grant PHY99-07949.
This work was partially supported by NASA grants NAG5-10825 and 
NNG04G177G and Texas Advanced Research Program grant 3658-0624-1999 to PRS.
The final draft of this paper was completed during visits by ES to CITA and of
PRS and ITI to the KITP at UCSB as part of the 2004 Galaxy-IGM Interactions 
Program.

\end{document}